\documentstyle[preprint,aps, eqsecnum]{revtex}
\tightenlines

\def\epsfig#1#2#3#4
         {
         \epsfysize=#2 \vbox{ \hglue#3 \epsfbox[#4]{#1} }
         }
\def\epsfigrot#1#2#3#4
         {
         \epsfxsize=#2
         \setbox\rotbox=\hbox to #2{\epsfbox[#4]{#1}}
         \vbox{\hglue#3 \rotl\rotbox}
         }
\newbox\rotbox
\begin{document}
\draft
\title{$gl(N|N)$ Super-Current Algebras for Disordered  Dirac Fermions
in Two Dimensions }
\author{S. Guruswamy$^a$, A. LeClair$^b$, A.W.W. Ludwig$^c$}
\address{$^a$ Institute for Theoretical Physics, 
Valckenierstraat 65, 1018 XE Amsterdam,}
\address{ THE NETHERLANDS }
\address{$^b$ Newman Laboratory, Cornell University, Ithaca, NY
14853.}
\address{$^c$ Department of Physics, University of 
California, Santa Barbara, CA 93106.}
\date{\today}
\maketitle

\begin{abstract}

We consider the  non-hermitian
2D  Dirac Hamiltonian with \underbar{\it (A):}   real  random  
mass,   imaginary scalar potential and  imaginary
gauge field potentials, and \underbar{\it (B):} arbitrary complex random
potentials of all three kinds. In both cases this Hamiltonian gives rise 
to a delocalization transition at zero energy with particle-hole
symmetry in every realization of disorder.
\underbar{\it Case (A)} is in addition time-reversal invariant,
and can also be interpreted as  
the random-field XY Statistical Mechanics model in two
dimensions.
The supersymmetric approach to disorder averaging  results 
in current-current perturbations of $gl(N|N)$ super-current algebras.  
Special properties of the $gl(N|N)$ algebra allow  the
exact  computation of the $\beta {\rm eta}$-functions,
and of the correlation functions of all  currents. One of them
is the Edwards-Anderson order parameter.
The theory is `nearly conformal' and
 possesses a scale-invariant subsector which is not
a current algebra. For $N=1$, in addition,  we obtain an  
exact solution
of all correlation functions.
We also study the  delocalization transition of \underbar{\it case (B)},  
with broken
time reversal symmetry, in the Gade-Wegner (Random-Flux)
 universality class,
using a $GL(N|N;C)/U(N|N)$ sigma model,
as well as
its $PSL(N|N)$ variant,  and
a corresponding generalized  random XY model.
For $N=1$ the sigma model  is shown to be identical to the
current-current perturbation.
For the
delocalization transitions (case (A) and (B))  a
density of states, diverging at zero
energy, is found. 

\vskip1cm
\noindent ITFA-99-15

\end{abstract}
\vskip 0.2cm


\newpage

\narrowtext
%
%
%
%
\def\oti{{\otimes}}
\def\bra#1{{\langle #1 |  }}
\def\lb{ \left[ }
\def\rb{ \right]  }
\def\tilde{\widetilde}
\def\bar{\overline}
\def\hat{\widehat}
\def\*{\star}
\def\[{\left[}
\def\]{\right]}
\def\({\left(}		\def\BL{\Bigr(}
\def\){\right)}		\def\BR{\Bigr)}
\def\<{\left\langle}  \def\>{\right\rangle} 

	\def\BBL{\lb}
	\def\BBR{\rb}
%
%
\def\zb{{\bar{z} }}
\def\zbar{{\bar{z} }}
\def\frac#1#2{{#1 \over #2}}
\def\inv#1{{1 \over #1}}
\def\half{{1 \over 2}}
\def\d{\partial}
\def\der#1{{\partial \over \partial #1}}
\def\dd#1#2{{\partial #1 \over \partial #2}}
\def\vev#1{\langle #1 \rangle}
\def\ket#1{ | #1 \rangle}
\def\rvac{\hbox{$\vert 0\rangle$}}
\def\lvac{\hbox{$\langle 0 \vert $}}
\def\2pi{\hbox{$2\pi i$}}
\def\e#1{{\rm e}^{^{\textstyle #1}}}
\def\grad#1{\,\nabla\!_{{#1}}\,}
\def\dsl{\raise.15ex\hbox{/}\kern-.57em\partial}
\def\Dsl{\,\raise.15ex\hbox{/}\mkern-.13.5mu D}
%
%
\def\th{\theta}		\def\Th{\Theta}
\def\ga{\gamma}		\def\Ga{\Gamma}
\def\be{\beta}
\def\al{\alpha}
\def\ep{\epsilon}
\def\vep{\varepsilon}
\def\la{\lambda}	\def\La{\Lambda}
\def\de{\delta}		\def\De{\Delta}
\def\om{\omega}		\def\Om{\Omega}
\def\sig{\sigma}	\def\Sig{\Sigma}
\def\vphi{\varphi}
%
%
\def\CA{{\cal A}}	\def\CB{{\cal B}}	\def\CC{{\cal C}}
\def\CD{{\cal D}}	\def\CE{{\cal E}}	\def\CF{{\cal F}}
\def\CG{{\cal G}}	\def\CH{{\cal H}}	\def\CI{{\cal J}}
\def\CJ{{\cal J}}	\def\CK{{\cal K}}	\def\CL{{\cal L}}
\def\CM{{\cal M}}	\def\CN{{\cal N}}	\def\CO{{\cal O}}
\def\CP{{\cal P}}	\def\CQ{{\cal Q}}	\def\CR{{\cal R}}
\def\CS{{\cal S}}	\def\CT{{\cal T}}	\def\CU{{\cal U}}
\def\CV{{\cal V}}	\def\CW{{\cal W}}	\def\CX{{\cal X}}
\def\CY{{\cal Y}}	\def\CZ{{\cal Z}}

\def\rvac{\hbox{$\vert 0\rangle$}}
\def\lvac{\hbox{$\langle 0 \vert $}}
\def\comm#1#2{ \BBL\ #1\ ,\ #2 \BBR }
\def\2pi{\hbox{$2\pi i$}}
\def\e#1{{\rm e}^{^{\textstyle #1}}}
\def\grad#1{\,\nabla\!_{{#1}}\,}
\def\dsl{\raise.15ex\hbox{/}\kern-.57em\partial}
\def\Dsl{\,\raise.15ex\hbox{/}\mkern-.13.5mu D}
%
%
%
\font\numbers=cmss12
\font\upright=cmu10 scaled\magstep1
\def\stroke{\vrule height8pt width0.4pt depth-0.1pt}
\def\topfleck{\vrule height8pt width0.5pt depth-5.9pt}
\def\botfleck{\vrule height2pt width0.5pt depth0.1pt}
\def\Zmath{\vcenter{\hbox{\numbers\rlap{\rlap{Z}\kern
0.8pt\topfleck}\kern 2.2pt
                   \rlap Z\kern 6pt\botfleck\kern 1pt}}}
\def\Qmath{\vcenter{\hbox{\upright\rlap{\rlap{Q}\kern
                   3.8pt\stroke}\phantom{Q}}}}
\def\Nmath{\vcenter{\hbox{\upright\rlap{I}\kern 1.7pt N}}}
\def\Cmath{\vcenter{\hbox{\upright\rlap{\rlap{C}\kern
                   3.8pt\stroke}\phantom{C}}}}
\def\Rmath{\vcenter{\hbox{\upright\rlap{I}\kern 1.7pt R}}}
\def\Z{\ifmmode\Zmath\else$\Zmath$\fi}
\def\Q{\ifmmode\Qmath\else$\Qmath$\fi}
\def\N{\ifmmode\Nmath\else$\Nmath$\fi}
\def\C{\ifmmode\Cmath\else$\Cmath$\fi}
\def\R{\ifmmode\Rmath\else$\Rmath$\fi}

\tableofcontents

\def\psib{\bar{\psi}}
\def\dtx{{\frac{d^2 x}{2\pi}}}
\def\dfx{{\frac{d^2 x}{4\pi}}}
\def\vep{\varepsilon}

\newpage

\load{\footnotesize}

\section{Introduction} 

Quantum field theory in the presence of quenched disorder is an increasingly
important subject in condensed matter physics.  Unfortunately, even in
two dimensions, fully
 solvable but non-trivial models are rather scarce. 
In this paper we consider, amongst others,  what   is perhaps the simplest
model of disordered fermions 
that can be exactly solved but also exhibits
some non-trivial behavior. We work in two spatial dimensions throughout.
 The model can be defined by the euclidean 
action\footnote{partition function $ Z=\int {\cal D}[\psi^\dagger,\psi]
\ \exp (- S)$}  

\begin{equation} 
\label{eq6}
S_0^f = \int \dtx   \Bigl(
\psi_{L}^\dagger ( \d_\zbar +  A_\zbar ) \psi_{L}
+ \psi_{R}^\dagger (\d_z +  A_z ) \psi_{R}
+ m(x)  ~  \psi_{R}^\dagger \psi_{L} + m^* (x)  ~  \psi_{L}^\dagger \psi_{R}
\Bigr) 
\end{equation}
where $d^2 x = dxdy$, $z=(x+iy)/2$, 
$\zbar = (x-iy)/2$,
and $A_z = A_x - i A_y$, $A_\zbar = A_x + i A_y $. 
The subscripts $L,R$ on the fermion fields denote left versus right movers
in the conformal theory with $m=m^* = 0$.  Letting $m=V+ iM, ~ m^* = V-iM$, 
we take $V$ and $M$ to have gaussian probability distributions of
equal\footnote{The difference of the variances of Gaussian random
variables $V$ and
$M$ renormalizes to zero.}
strength $g$ and zero mean, and similarly for $A_x , A_y$ with
strength $g_A$: 
\begin{eqnarray}
P[m,{m^*}] &=& \exp \( -\inv{g} \int
\dtx ~ {m^*} (x)  m(x)  \)
\\ \nonumber
P[A_z , A_\zbar] &=&
\exp \( -\inv{g_A} \int \dtx ~ A_z (x)  A_\zbar (x)  \)
\label{eIii}
\end{eqnarray}
(The couplings $g, g_A$ represent twice the variance.) 
Disorder averaging corresponds to a gaussian functional integral:
\begin{equation}
\label{eq8}
\bar{\langle \CO \rangle} = \int Dm^* Dm DA_z DA_\zbar ~
P[m,{m^*}] P[A_z , A_\zbar]
~ \langle \CO \rangle
\end{equation}
            
In the context of two-dimensional statistical mechanics, the above
model may be thought of as a  random field $XY$ model 
or equivalently the random field sine-Gordon model\cite{BernardrandomXY},   
at the free-fermion point. 
 The usual
sine-Gordon model has $m$ constant, which now acquires a random phase,
preventing the theory from becoming massive.  
Random field XY models were first studied by Cardy and 
Ostlund\cite{cardyostlund}
using replicas\cite{mudryXY}.
The action in Eq.(\ref{eq6})
can be related (see below)  to
a  2D Dirac hamiltonian $H_2$
subject to a {\it real}  random mass $M$,
a  random {\it imaginary}  potential $-iV$ with strength $g_M = g_V = g$, 
and also in a random {\it imaginary}  gauge-field
\begin{equation}
\label{eq1}
H_2 = (-i\d_x +i  A_x ) \sigma_x + 
(-i\d_y + i A_y ) \sigma_y + M(x,y)  \sigma_3
-i   V(x,y)
\end{equation}
where $\sigma_i$ are the standard Pauli matrices.
This hamiltonian is non-hermitian.
This theory should be compared with the models
described in Ref.'s \cite{LFSG},\cite{Chalker}  which  were argued
to have a transition in the universality class of the quantum Hall
plateau
transition;  the latter models   have a  {\it real} random mass 
and a  {\it real}  scalar  potential with strength   
$g_M \neq g_V$, and a {\it real} random gauge field,
the corresponding 2D Dirac hamiltonian being 
hermitian\footnote{We point out
that for this Quantum Hall setting, 
based on the $\beta$eta functions computed in \cite{Bernard}, 
 the line $g_M + g_V = 0$ which is the subject of this paper, 
is a line of fixed points in the ultraviolet. 
One can check numerically that starting from 
$g_{M }$, $g_{V}$, $g_{A}$ all positive,
one can actually flow to the line $g_M + g_V=0$ in the UV.}.  
The same model as in Eq.(\ref{eq1}),
also describes\footnote{We were first made aware of the relationship
between the model of Hatsugai et al. and the random $XY$ model
 by M.P.A. Fisher\cite{mpaf} in the context of the replica field theory.}
 the continuum limit of electrons
hopping on a square lattice with $\pi$ flux\cite{hatsugai},
exhibiting  a delocalization transition at zero energy,
first discussed by Hatsugai, Wen and Kohmoto[HWK] in Ref. \cite{hatsugai}.
This involves a doubling of the number of degrees of freedom( see  section II
and Appendix B), leading to a  random {\it hermitian} 4-component
hamiltonian.  Applications of our
results to this  delocalization  transition 
will be described in section $V$.

We perform the disorder averaging using the supersymmetric method.
This leads to a left-right current-current perturbation 
of the $gl(N|N)$ conformal supercurrent algebra for the $N$-species
version, used to compute $N$th moment disorder  averages. 
Conformal field theory methods were previously used by
Bernard for the case  where $g_M ({\rm or}~ g_V) = 0$\cite{Bernard},
which leads to the $osp(2N|2N)$ algebra.  Also, the  
case of $g_V=g_M = 0$, but $g_A \neq 0$, where the conformal
symmetry is unbroken and the model is equivalent to free bosons,  
was studied in 
\cite{Wen},  \cite{LFSG}. 
Though our model has a  non-zero renormalization group 
$\beta {\rm eta}$-function,
some special features of the $gl(N|N)$ algebra, in particular the
nilpotency of the fermionic generators of $gl(N|N)$ and the 
existence of two quadratic casimirs, lead to a complete solution 
for {\it all} correlation functions in the case $N=1$,
and  to the solution of the correlation functions of
all Noether currents in the case of general $N$.  
The only non-vanishing
$\beta$etafunction is that for
$g_A$ which we compute exactly. It only depends on $g$.
Integration of the  renormalization group
equations allows us to compute for instance the density of states
of the delocalization problem of \cite{hatsugai}.  

Our results reveal some interesting features of the current-current
perturbations of Lie superalgebras in comparison with the analogous
perturbation of the ordinary bosonic Lie algebras\footnote{
The current-current perturbation of $su(2)$ at level $1$ gives the
Gross-Neveu model, which is a massive theory with factorizable
S-matrix.  See for instance \cite{BL}. }.   
Our models are integrable quantum field theories for
the usual reasons (Lax pair, etc.), and are not conformal.  However
the structure of the theory allows a solution which doesn't rely on
exact S-matrices, form-factors, etc.  Regarding the difficulty of
solution they lie somewhere between the conformal field theories
and the massive integrable theories. 

Delocalization transitions of `sublattice' (or `random flux') models,
 studied first by Hikami
et al. \cite{HikamiShWeg}, and by Gade and Wegner\cite{Gade}
perturbatively using replica sigma models, have been known for
some time to exhibit
R.G. beta functions of the kind we obtained for the current-current
perturbations of $gl(N|N)$ supercurrent algebras. These delocalization
transitions possess a special particle-hole symmetry, known to
prevent localization due to disorder even in one dimension\cite{McCoyWu}.
This is not an accident:
In the last section of this paper we exhibit a random
Dirac Hamiltonian giving rise to these theories (in 2D).
{}From this Dirac Hamiltonian we derive a SUSY sigma model, which
we find to live on a target manifold $GL(N|N;C)/U(N|N)$.
We solve this sigma model exactly for the case $N=1$, 
in the presence of
a topological Wess-Zumino-Witten(WZW) term
with arbitrary real coupling $k$.
 We find that the $GL(1|1;C)/U(1|1)$
sigma model with WZW term is the same theory as 
the 
current-current
perturbation of $gl(1|1)$ supercurrent algebra.
One expects that such an
 equivalence should extent to the case of general $N$.

An interesting and  special feature of our current-current perturbations is
that (for any value of $N$)  the non-vanishing
$\beta$etafunction is associated only with two of the  generators
of the $gl(N|N)$ algebra. These can be separated out
and a scale invariant fixed point theory results for any value of
the disorder strength $g$, i.e. a line of fixed points (for any
value of
$N$).  
After we obtained the current algebra results in section  IV,
the papers \cite{Bershadsky}\cite{Witten} appeared on
$PSL(N|N)$ sigma models, exhibiting a line of fixed points
as the sigma  model coupling constant is varied.
We have added
a paragraph to 
section IV C explaining the connection of our results with
these papers.
 Our bosonized formulation may provide
a useful way of studying the $PSL(N|N)$ sigma models.
Also, a proposal for a relationship between the
$PSL$ sigma models and the integer quantum Hall plateau
transition has appeared recently\cite{Zirnbauer}.

The outline of the paper is as follows. In 
section  II we give a general symmetry-based overview
of the models we discuss in this paper.
Sections III and IV  address the
random field XY model.
In section III 
we formulate the supersymmetric method and extend it
to the $N$-species version.  In section IV A we compute the
 $\beta {\rm eta}$-function for the random field XY model
 to all orders, and the exact
 correlation functions  
 using current algebra techniques. In section IV B
some of these results are extended to the $N$-species case,
possessing $gl(N|N)$ symmetry.
The scale invariant subsector of this theory is  discussed
in more detail in Section IV C.
A generalized random field XY model,  whose associated Dirac
hamiltonian lacks time-reversal
symmetry is discussed in Section IV D. (This Dirac Hamiltonian
underlies the Gade-Wegner `sublattice' models discussed 
later in section VI.)
In section V we turn to delocalization transition of 
Hatsugai et al.\cite{hatsugai} 
and compute the density
of states as a function of energy. 
 In section  VI A we discuss the corresponding delocalization
transition with {\it broken time reversal} symmetry,
which is in the Gade-Wegner universality class\cite{Gade} 
and find that it is described by a $GL(1|1;C)/U(1|1)$
sigma model, which we solve exactly. In section VI C we relate
the  sigma model
to the current-current perturbation  discussed earlier  (in section IV A).
In section VI C we conclude with a summary of the $N$-species generalization,
the $GL(N|N;C)/U(N|N)$ sigma model, and its scale invariant $PSL(N|N)$ variant.
A number of technical details are delegated to  three Appendices.
In Appendix A we relate the $N=1$ species theory
discussed in  
section IV A,  to the random field XY model.
Appendix B   gives the steps leading from the general
non-hermitian Dirac hamiltonian and
its hermitianization 
to a corresponding conformal field theory action.
Finally,  Appendix C contains details of the 1-loop RG equations 
of Section  IV D.

\section{ The Models and their Symmetries: An Overview}

The quantum mechanics of non-interacting particles (electrons) subject to
random potentials and their   possible
universality classes of localization/delocalization transitions
are conventionally classified according to 
those symmetries which are present for
a fixed realization of disorder. In this section we describe
the class of models  discussed 
in this paper, all of which exhibit
particle-hole symmetry.

\subsection{Time Reversal Symmetric Models}

The fundamental model that we study in this
paper can be formulated as the random field XY model\cite{cardyostlund}.
This is a 2D statistical mechanics model.
At the free fermion point of the XY model, however,  the XY spin operator
can be  represented, using abelian bosonization, by
a bilinear $\psi^\dagger_L\psi_R$  of Dirac fermions
governed by the action of Eq.(\ref{eq6}).
Alternatively, this action
 can also  be related to
the 2-component random  Dirac Hamiltonian $H_2$ of Eq.(\ref{eq1}) in
two spatial dimensions, 
containing a  random
real Dirac mass $M$, a random imaginary scalar potential
$V$ and purely imaginary gauge potential ${\vec A}$ term.
(Details of this connection are lain out in Appendix A). 
The latter being a non-interacting system, 
this allows us to use the supersymmetry (SUSY) method
for disorder averaging. This is the method we use
in this paper to study the random XY models.
An apparent difficulty in applying
this method  is the  {\it non-hermiticity}\cite{nonhermQM}
of this quantum mechanical  hamiltonian  $H_2$, which seems to
prevent the functional integral over the bosonic fields,
appearing in the
  SUSY approach to disorder,  from converging. This problem is
however easily remedied by considering an
associated hermitian 4-component hamiltonian (Appendix A)
\begin{equation}
\label{Hfourintro}
H_4
\equiv
\pmatrix{ 0 & H_2 \cr
          H_2^\dagger & 0\cr}
\end{equation}
As discussed by HWK \cite{hatsugai}, the hamiltonian $H_4$ itself
describes
the quantum mechanics (in the continuum limit)
of  fermions hopping, with real hopping
amplitudes,  on a 2D square lattice with
$\pi$-flux through each plaquette (see also \cite{fukui}). This theory 
 possesses  a delocalization transition at zero energy,
the characteristic symmetries of which are 
particle-hole and time-reversal  symmetries\footnote{
For the action of these symmetries in the continuum theory
with hamiltonian $H_4$, see Eq. (\ref{Cparticlehole},\ref{Treversal}).}.
Hence there are two apparently unrelated problems
linked to this 4-component hamiltonian, the random
field XY model, and  the lattice fermion hopping model.
In the latter model, a small imaginary part has
to be added to define the quantum mechanical Green's
functions. This is not necessary for the interpretation
as an random XY statistical mechanics model.
As discussed in more detail below (and in Appendix A),  
the theory describing averaged quantities of
the  random XY model has $gl(1|1)$ global SUSY,
and $gl(N|N)$ SUSY  when higher moment averages
(e.g.  of `Edwards-Anderson order parameter'
type) are considered.
The corresponding  delocalization transition\cite{hatsugai}, on the
other hand,  can be shown to be a $GL(2|2;R)/Osp(2|2)$
 sigma model.

\subsection{Models with  Broken Time Reversal Symmetry}

In the random field XY model discussed above, 
the complex XY spin, represented by 
$\psi^\dagger_L\psi_R$ 
is coupled to a random ordering field,
which is  a random phase $m(x)=m e^{i\vartheta}$
with $\vartheta$ real.
The corresponding random 2D Dirac Hamiltonian $H_4$ $\bigl ($ of  
Eq. (\ref{eq1}, \ref{Hfourintro}) $\bigr )$
is time reversal invariant.
Time reversal symmetry  breaking  may for
example be introduced by  allowing, 
in the language of the Dirac Hamiltonian $H_2$,  
in addition to the  purely {\it imaginary}  gauge potential discussed above,  
also
a  {\it real}  gauge potential\footnote{
Properties of 
the most general Dirac hamiltonian where random Dirac mass, scalar
and gauge potentials have both real and imaginary parts, as well
as the relationship with a corresponding SUSY field theory, are
summarized in Appendix B.}.
The addition of a {\it real} random gauge potential
leads to the generation of both, real and imaginary parts
of all three types of potentials,
and the theory flows off to strong coupling\footnote{
 See Eq.(\ref{RGnotimerev})
of section IV D, with the notations of Appendix B}.
It is then more useful to employ a description 
in terms of a non-linear sigma model (section VI).
The corresponding quantum mechanical hamiltonian
$H_4$, possessing particle-hole symmetry but not time-reversal
symmetry,  is now, by universality,   in the class of the
Gade-Wegner sublattice models\cite{Gade},\cite{Altland},
describing fermions hopping 
on the square lattice with {\it complex} hopping amplitudes.
The corresponding random XY model has $GL(1|1;C)$ global
SUSY. In section VI we describe the 
corresponding Gade-Wegner type  delocalization transition 
in terms of a non-linear sigma model on
a manifold which we denote by
$GL(1|1;C)/U(1|1)$
(allowing also for a topological Wess-Zumino term.)

It is noteworthy to re-emphasize that both models, 
with or without time reversal symmetry,
 possess particle-hole symmetry at zero
energy,  a feature
which  can  circumvent localization of all states
due to disorder, at that energy, even in one dimension\cite{McCoyWu}.

\section{Supersymmetric Disorder Averaging} 

In this section and the following sections IV A, B, C,  
we discuss the random field XY model,
 described by the  hamiltonian of Eq.(\ref{eq1}),
 which we can solve exactly.

To implement the supersymmetric method for disorder averaging\cite{Efetov}, 
we augment the theory with
bosonic ghosts $\beta^\dagger_{L,R}, \beta_{L,R}$, 
coupled to the disorder in the
same way as the fermions, with action $S_0^b = S_0^f (\psi \to \beta)$.
The partition function of the fermions plus ghosts is now independent of
the disordered potentials\footnote{ There is a subtlety
arising from  the convergence of the bosonic functional
integral.  A careful discussion of this issue, done in Appendix A,
results in the effective action described below.}
so the order of functional integration
over the `matter-ghosts'  and disorder can be interchanged.  Integrating
over disorder, this yields an effective action for the fermions and ghosts:
\begin{equation}
\label{eq11}
S_{\rm eff} = \int \dtx \(   \( \psi_{L}^\dagger \d_\zbar
\psi_{L} + \psi_{R}^\dagger \d_z \psi_{R}  + \beta_{L}^\dagger
\d_\zbar \beta_{L} + \beta_{R}^\dagger \d_z \beta_{R} \)
+ g \CO_m + g_A \CO_A   \)
\end{equation}
where
\begin{eqnarray}
\CO_m  &=& -  \( \psi_{R}^\dagger \psi_{L} + \beta_{R}^\dagger
\beta_{L} \) \( \psi_{L}^\dagger \psi_{R} + \beta_{L}^\dagger \beta_{R}
\)
\\  \nonumber
\CO_A &=& - \( \psi_{L}^\dagger \psi_{L} +
\beta_{L}^\dagger \beta_{L} \) \( \psi_{R}^\dagger \psi_{R}
+ \beta_{R}^\dagger \beta_{R} \)
\label{eq13}
\end{eqnarray}

In the sequel we will need to introduce a multi-species version of
the above model.  
Introducing a species index $\psi_{La}$, $a=1, .., N$,
we define the $N$-species
model as
\begin{equation}
S_0^f = \int \dtx  \sum_{a=1}^N \Bigl(
\psi_{La}^\dagger ( \d_\zbar +  A_\zbar ) \psi_{La}
+ \psi_{Ra}^\dagger (\d_z +  A_z ) \psi_{Ra}
+ m(x) \psi_{Ra}^\dagger \psi_{La} + {m^*} (x) \psi_{La}^\dagger \psi_{Ra}
\label{eq6b}
\end{equation}
Introducing bosonic ghosts and integrating over disorder leads to the
effective action:
\begin{equation}
\label{eq11N}
S_{\rm eff}^{(N)} = \int \dtx \(  \sum_{a=1} \( \psi_{La}^\dagger \d_\zbar
\psi_{La} + \psi_{Ra}^\dagger \d_z \psi_{Ra}  + \beta_{La}^\dagger
\d_\zbar \beta_{La} + \beta_{Ra}^\dagger \d_z \beta_{Ra} \)
+ g \CO_m + g_A \CO_A  \)
\end{equation}
where now
\begin{eqnarray}
\CO_m  &=& - \sum_{a,b} \( \psi_{Ra}^\dagger \psi_{La} + \beta_{Ra}^\dagger
\beta_{La} \) \( \psi_{Lb}^\dagger \psi_{Rb} + \beta_{Lb}^\dagger \beta_{Rb}
\)
\\
\CO_A &=& - \sum_{a,b} \( \psi_{La}^\dagger \psi_{La} +
\beta_{La}^\dagger \beta_{La} \) \( \psi_{Rb}^\dagger \psi_{Rb}
+ \beta_{Rb}^\dagger \beta_{Rb} \)
\label{eq13N}
\end{eqnarray}

\section{ Random XY Models via $gl(N|N)$ Current Algebras}

\subsection{ Random Field XY Model: $N=1$ Species}

In this section we consider the $N=1$ model in detail.    
The bosonic ghosts $\beta_L $ ($\beta_R$) have Lorentz spin
$1/2$ ($-1/2$).  The conformal field theory for such a bosonic
first order action was treated in ref. \cite{FMS}.  The Virasoro
central charge for one copy of the $\beta^\dagger , \beta$ system
is $c=-1$, so that the total central charge of the matter plus ghosts
is zero.  This had to be the case since the partition function is one.

In the effective action eq. (\ref{eq11}) for $N=1$ species 
the operators $\CO_m$ and $\CO_A$ are bilinears in the dimension $1$
currents:
\begin{eqnarray}
\label{eIv}
G_+ = \beta_L^\dagger  \psi_L  ,~~~~~ G_- = \beta_L  \psi_L^\dagger 
,~~~~~ J= :\psi_L  \psi_L^\dagger :, ~~~~
J' = :\beta_L \beta_L^\dagger: 
\\
\bar{G}_+ = \beta_R^\dagger \psi_R, ~~~~~ \bar{G}_- = \beta_R \psi_R^\dagger,
~~~~~ \bar{J} = : \psi_R \psi_R^\dagger :, ~~~~~ \bar{J}' = :
\beta_R \beta_R^\dagger : 
\end{eqnarray}
(We denote all left-right currents as $(J_L , J_R ) = (J, \bar{J} )$.) 
Namely, 
\begin{eqnarray}
\CO_m &=&  J\bar{J} - J' \bar{J}' - G_- \bar{G}_+ + G_+ \bar{G}_- 
\\  \nonumber
\CO_A &=& - ( \bar{J}' - \bar{J} )(J'-J) 
\label{eIvi}
\end{eqnarray} 
 
\def\gl{$gl(1|1)$}

In the conformal limit $g=g_A = 0$, the currents generate a
\gl ~ current algebra. The currents $J, J'$ are bosonic $U(1)$ currents
for the fermions and ghosts respectively, and $G_\pm$ are their
fermionic superpartners.  {}From the operator product expansions (OPE)
\begin{equation}
\psi_L (z) \psi_L^\dagger (w) =  \psi_L^\dagger (z) \psi_L (w) \sim \inv{z-w} , 
~~~~~
\beta^\dagger_L  (z) \beta_L (w) =  - \beta_L (z) \beta_L^\dagger
 (w) \sim -\inv{z-w}   
\end{equation}
one obtains the OPE's:
\begin{eqnarray}
J(z) J(w) &\sim& \inv{(z-w)^2} , ~~~~~~~ J' (z) J' (w) \sim - \inv{(z-w)^2} 
\\ \nonumber 
J(z) G_\pm (w) &\sim& \pm \inv{z-w} G_\pm (w), ~~~~~ 
J'(z) G_\pm (w) \sim \pm \inv{z-w} G_\pm (w)
\\ \nonumber
G_- (z) G_+ (w) &\sim& \inv{(z-w)^2} +\inv{z-w} \( J(w) - J' (w) \) 
\label{eIvii}
\end{eqnarray}

Let $j, j', g_\pm$ denote the zero modes of the currents, defined as
$j = \oint \frac{dz}{2 \pi i} J(z)$, etc. 
These zero modes are generators for the global \gl algebra\footnote{
For the complete set of modes the current algebra (\ref{eIvii})
defines the affine Lie superalgebra $\hat{gl(1|1)}$.}:
\begin{equation}
\label{eIviii}
[j, g_\pm] = [j' , g_\pm ] = \pm g_\pm , ~~~~~\{ g_+ , g_- \} = j-j' 
, ~~~~~g_\pm^2 = 0
\end{equation}
The algebra \gl has two quadratic Casimirs:
\begin{equation}
\label{eIix}
C = j^2 - j'^2 - g_- g_+ + g_+ g_- , ~~~~~
\tilde{C} = (j-j')^2 
\end{equation}
The operators $\CO_m$ and $\CO_A$ are left-right current-current
perturbations with precisely the structure of the Casimirs $C, \tilde{C}$
respectively.  This implies that the $gl(1|1)_L \otimes gl(1|1)_R$
symmetry of the conformal field theory is broken to a diagonal 
\gl symmetry in $S_{\rm eff}$.  

The model defined by $S_{\rm eff}$  is a perturbation of a $c=0$
conformal field theory, and can be studied in the framework of
Zamolodchikov\cite{Zamo}.  The $gl(1|1)$ conformal field theory
was studied in \cite{Saleur}.    Current-current perturbations 
are generically integrable in the case of the bosonic Lie algebras, 
which means there are an infinite number of conserved currents. 
It is likely that this feature also holds for the current-current
perturbations of the superalgebras.  
 However we do not pursue this
here since, as we now show, the model can be solved by straightforward
manipulations of the functional integral.

The manner in which conformal symmetry is broken in $S_{\rm eff}$ 
is contained in the $\beta{\rm eta}$-functions.  The OPE's of the
perturbing operators are the following: 
\begin{equation}
\label{eIx}
\CO_m (z,\zbar) \CO_m (0) \sim \frac{- 2}{z\zbar} \CO_A (0), ~~~~~~
\CO_A (z,\zbar) \CO_m (0) \sim 0, ~~~~~ 
\CO_A (z,\zbar) \CO_A (0) \sim 0 
\end{equation}
Note that because of the supersymmetry, the $1/(z\zbar)^2 $ terms vanish. 
One obtains to lowest order \cite{ZamoRG} 
\begin{eqnarray}
\beta_A (g) &=& \frac{d g_A}{d l} =  2 g^2 +...
\\ \nonumber
\beta_g  &=& 0  
\label{eIxi}
\end{eqnarray}
where $l$ is the $log$ of the rescaling factor.
 The ultraviolet limit corresponds
to $l\to -\infty $ where $g_A \to -\infty$, whereas the infrared limit
is $l \to \infty$  with $g_A \to +\infty$.   
This shows that if $g_A$ disorder were not included from the start
it would be generated under renormalization. 
Since the operator $\CO_m$ is never generated in the OPE at higher
orders, it follows that $\beta_g = 0$ to all orders.  

\def\vphi{\varphi}
\def\vphib{{\bar{\varphi}}}

It will be useful to bosonize the theory.  The $U(1)$ fermion current
is bosonized with a scalar field $\phi$: 
\begin{equation}
\label{eIxii}
J = i \d_z \phi , ~~~~~ \bar{J} = -i \d_\zbar \phi 
\end{equation}
In the conformal limit, $\phi (z, \zbar) = \varphi_L (z) + 
\varphi_R (\zbar) $ and 
\begin{equation}
\label{eIxiii}
\psi_L = \exp ({ i \varphi_L }) , ~~\psi_L^\dagger = \exp( -i\varphi_L ), 
 ~~~~~\psi_R  = \exp ({- i \varphi_R} ), ~~ \psi_R^\dagger = 
\exp ( i \varphi_R ) 
\end{equation}

Bosonization of the $\beta^\dagger-\beta$ system was described in \cite{FMS}.
The $J'$ currents are expressed in terms of another scalar field
$\phi'$:
\begin{equation}
\label{eIxiv}
J' = i \d_z \phi' , ~~~~~ \bar{J}' = -i\d_\zbar \phi' 
\end{equation}
The scalar field $\phi'$ has the opposite sign kinetic term compared
to $\phi$, which implies:
\begin{equation} 
\label{eIxivb}
\phi (z, \zbar) \phi (0 ) \sim - \log z\zbar , ~~~~~~
\phi' (z, \zbar) \phi' (0 ) \sim + \log z\zbar 
\end{equation}
Since the $\phi'$ field still contributes $c=1$,  in order to obtain 
the $c=-1$ of the $\beta^\dagger -\beta$ system one needs an additional
fermionic $\eta-\xi$ system with $c=-2$.  The result is 
\begin{equation}
\label{eIxv}
\beta_L^\dagger  = e^{-i\vphi'_L} \d_z \xi_L, ~~~~~ \beta_L = 
e^{i\vphi'_L} \eta_L , 
~~~~~
\beta^\dagger_R  = e^{i\vphi'_R} \d_\zbar \xi_R , ~~~~~
\beta_R  = e^{-i \vphi'_R } \eta_R 
\end{equation}
where as before $\phi' = \vphi'_L  + \vphi'_R$. 

The $\eta-\xi$ system has the first order action 
\begin{equation}
\label{eIxvi}
S_{\eta-\xi} = \int \dtx ~  \eta_L  \d_\zbar \xi_L + \eta_R  \d_z \xi_R 
\end{equation}
In order to have a more symmetric effective action it is convenient 
to formulate the $\eta-\xi$ system in terms of a complex 
dimension zero fermion $\chi$, with the action\cite{Kausch} 
\begin{equation}
\label{eIxvii}
S_\chi = \int \dfx  \( \d_z \chi^\dagger \d_\zbar \chi + 
\d_\zbar \chi^\dagger \d_z \chi \) 
\end{equation}
The equation of motion $\d_z \d_\zbar \chi = 0$ implies that 
$\chi $ separates into left and right moving parts:
$\chi = \chi_L (z) + \chi_R (\zbar)$.  The $\eta-\xi$ fields are related
to $\chi$ as follows:
\begin{eqnarray}
\d_z \xi_L  &=& i \d_z \chi , ~~~~~ \d_\zbar \xi_R = i \d_\zbar \chi 
\\ \nonumber
\eta_L  &=& i \d_z \chi^\dagger , ~~~~~ \eta_R = i \d_\zbar \chi^\dagger
\label{eIxviii}
\end{eqnarray}

In order to absorb the $-J\bar{J} + J' \bar{J}'$ terms in $\CO_m$ 
into the kinetic energy, we define
\begin{equation}
\label{eIxix}
\phi_\pm = \inv{\zeta} \( \phi \pm \phi' \) , ~~~~~~
\zeta \equiv \inv{\sqrt{1+2g}} 
\end{equation}
Putting all of this together one obtains the 
effective action
\begin{eqnarray}
S_{\rm eff} &=& \int \dfx  
\Bigl[  
\d_z \phi_+ \d_\zbar \phi_-  
- 2 \zeta^2 g_A \d_z \phi_- \d_\zbar \phi_- 
+ \( 1 + 2g e^{-i\zeta \phi_-} \) \d_z \chi^\dagger \d_\zbar \chi 
\\ \nonumber
&~&~~~~~~~~~
+ \( 1 + 2g e^{i\zeta \phi_-} \) \d_\zbar \chi^\dagger \d_z \chi 
\Bigr] 
\label{Seff}
\end{eqnarray}

\def\bfPhi{{\bf \Phi}}

In order to compute correlation functions, 
we treat the effective theory in a number of steps.  First note
that 
\begin{equation}
\label{eIxx}
S_{\rm eff} \[ \phi_+, \phi_- , g_A \] 
= S_{\rm eff} \[ \phi_+ - 2 \zeta^2 g_A \phi_- , \phi_- , g_A=0 \] 
\end{equation}
Let $\bfPhi( \phi_+, \phi_-, \chi )$ be a composite field or product
of such fields at different space-time points.  Making a change of
variables $\phi_+ \to \phi_+ + 2\zeta^2 g_A \phi_-$ in the functional
integral, one relates correlation functions with $g_A \neq 0$ to 
those in the $g_A = 0$ theory: 
\begin{equation}
\label{eIxxi}
\langle \bfPhi (\phi_+ , \phi_- , \chi ) \rangle_{g_A} 
= \langle \bfPhi(\phi_+ + 2 \zeta^2 g_A \phi_- , \phi_- , \chi ) 
\rangle_{g_A = 0} 
\end{equation}
Henceforth $S_{\rm eff}$ denotes the effective action (\ref{Seff})
with $g_A = 0$.  

The next step recognizes that $\phi_+$ acts as a Lagrange multiplier. 
Now let $\bfPhi(\phi_- , \chi)$ denote a field or product of fields
that is independent of $\phi_+$.  Introducing a source $\rho$ for $\phi_+$
then functionally integrating over $\phi_+$ gives a functional 
$\delta$ function:
one has
\begin{eqnarray}
\< e^{-\int \dfx \rho \phi_+ } \bfPhi (\phi_- , \chi ) \>_{g_A = 0}
&=&
\int D[\phi_-, \chi ]  \delta (\d_z \d_\zbar \phi_- - \rho ) 
e^{-S_{\rm eff} [\phi_+ = 0] } \bfPhi ( \phi_- , \chi ) 
\\ \nonumber
&=& \int D[\chi^\dagger , \chi ] 
e^{-S_{\rm eff} [ \phi_+ = 0 , \phi_- = \hat{\rho} ]} 
\bfPhi( \phi_- = \hat{\rho} , \chi ) 
\label{eIxxii}
\end{eqnarray}
where
$\hat{\rho}$ is the solution to 
\begin{equation}
\label{eIxxiii}
\d_z \d_\zbar \hat{\rho} = \rho 
\end{equation}
(We have set the partition function to $1$.) 

\def\rhoh{{\hat{\rho}}}

Insertions of the operator $\d_z \d_\zbar \phi_+$ in a correlation
function can be obtained from functional derivatives with respect to
$\rhoh$. 
For instance, 
\begin{equation}
\label{eIxxiv}
\< \d_z \d_\zbar \phi_+ (z , \zbar ) \bfPhi (\phi_- , \chi ) 
\>_{g_A = 0} 
= -4\pi \[ \frac{\delta}{\delta \rhoh (z,\zb)} 
\int D\chi e^{-S_{\rm eff} [\phi_+ = 0, \phi_- = \rhoh]} 
\bfPhi (\phi_- = \rhoh , \chi )  \]_{\rhoh = 0} 
\end{equation}  
One needs 
\begin{equation}
\label{eIxxv}
4\pi \frac{\delta}{\delta\rhoh} 
S_{\rm eff} [\phi_+ = 0, \phi_- = \rhoh] \Bigr|_{\rhoh = 0} 
= 2ig\zeta \( \d_z \chi^\dagger \d_\zbar \chi - \d_\zbar \chi^\dagger
\d_z \chi \) 
\end{equation}
and 
\begin{equation}
\label{eIxxvi}
S_{\rm eff} [\phi_+ = \phi_- = 0] = (1+2g) S_\chi  
\end{equation}
We define a rescaled $\chi$ field,
\begin{equation}
\label{eIxxvii}
\hat{\chi} = \inv{\zeta} \chi 
\end{equation}
such that 
\begin{equation}
\label{rescale}
S_{\hat{\chi}} = (1+2g) S_{\chi}  
= 
\int \dfx  \( \d_z \hat{\chi}^\dagger \d_\zbar \hat{\chi} +
\d_\zbar \hat{\chi}^\dagger \d_z \hat{\chi} \)
\end{equation}

\def\schi{{ \hat{\chi}}}

The above formulas imply that all correlation functions are reduced
to a free-field correlators in the rescaled $\chi$ system defined
by $S_{\hat{\chi}}$.  
For instance:
\begin{eqnarray} 
\label{eIxxviii}
\< \d_z \d_\zbar \phi_+ (z , \zbar ) \bfPhi (\phi_- , \chi )
\>_{g_A = 0}
&=&
2ig\zeta \< \( \d_z \chi^\dagger \d_\zbar \chi - \d_\zbar \chi^\dagger
\d_z \chi \) \bfPhi(\phi_- =0, \chi )\>_\schi  
\\ 
&~& ~~~~~~~~~
- 4\pi \< \[ \frac{\delta \bfPhi (\phi_- ,\chi)}{\delta \phi_- (z,\zbar)} 
\]_{\phi_- = 0}  \>_{\schi} 
\end{eqnarray}
All correlation functions $\langle ~~\rangle_{\schi}$ on the RHS of the above
equation are computed with the free action $\schi$ (\ref{eIxvii}).
In particular
\begin{equation}
\label{chiprop}
\< \schi^\dagger (z,\zbar) \schi (0) \> = 
- \< \schi (z,\zbar) \schi^\dagger (0) \> = -\log (z\zbar) 
\end{equation}

The second term in eq. (\ref{eIxxviii})  
generally involves $\delta$ functions.
More generally the insertion of $\d_z \d_\zbar \phi_+$ at different
space-time points is equivalent to the insertion of the operator 
on the RHS of eq. (\ref{eIxxv})
at the same points in the free theory $\schi$
up to $\delta$-functions:
\begin{equation}
\label{eIxxix}
\< \( \d_z \d_\zbar \phi_+ \)^n \bfPhi (\phi_- \chi ) \>_{g_A =0} 
= (2ig\zeta)^n 
\< \( \d_z \chi^\dagger \d_\zbar \chi - \d_\zbar \chi^\dagger \d_z \chi
\)^n \bfPhi (\phi_- = 0 , \chi ) \>_{\schi} 
+ \delta -{\rm functions} 
\end{equation} 

The above formulas are sufficient to compute all correlation functions.
First, for correlations not involving $\phi_+$, the field $\phi_-$ may be
set to zero:
\begin{equation}
\label{eIxxxiv}
\< {\bf \Phi} (\phi_- , \chi ) \>_{g_A = 0} 
= \< {\bf \Phi} (\phi_- = 0, \chi ) \>_{g_A = 0} 
\end{equation} 
Thus,
\begin{equation}
\label{eIxxxv}
\< \phi_- (z,\zbar) \phi_- (0) \> = 0
\end{equation}
Equation (\ref{eIxxviii}) implies:
\begin{equation}
\label{eIxxxvi}
\d_z \d_\zbar \< \phi_+ (z,\zbar) \phi_- \>_{g_A = 0}  = -4\pi \delta^2 (x) 
\end{equation} 
Thus to all orders in $g$:
\begin{equation}
\label{eIxxxvii}
\< \phi_+ (z,\zbar) \phi_- (0) \>_{g_A = 0} = -2 \log (z\zbar) 
\end{equation} 

Correlation functions involving multiple $\phi_+$ insertions are
more interesting.  {}From (\ref{eIxxix}) one finds
\begin{equation} 
\label{eIxxxviii}
\< \d_z \d_\zbar \phi_+ (z,\zbar) \d_w \d_{{\bar{w}}} \phi_+ (w,\bar{w} ) 
\> = -8g^2 \zeta^6  \inv{ (z-w)^2 (\zbar - \bar{w})^2 } 
+ \delta-{\rm functions} 
\end{equation} 
Integrating this, one finds a  $\log^2$ contribution: 
\begin{equation}
\label{eIxxxix}
\< \phi_+ (z,\zbar) \phi_+ (0) \>_{g_A = 0 } =
 -4 g^2 \zeta^6 \log^2 (z\zbar /a^2 )
\end{equation} 
where $a$ is an ultraviolet cutoff scale.

Finally let us compute correlation functions of the ` matter' $U(1)$ field
$\phi = \zeta (\phi_+ + \phi_- )/2 $ with $g_A \neq 0$.  Using 
eq. (\ref{eIxxi})  one obtains
\begin{equation} 
\label{eIxxxx}
\< \phi (z,\zbar) \phi (0) \> = -\zeta^2 (1+2\zeta^2 g_A ) \log(z\zbar/a^2) 
- g^2 \zeta^8 \log^2 (z\zbar/a^2 ) 
\end{equation} 
Similarly,
\begin{equation}
\label{eIxxxxb}
\< \phi' (z,\zbar) \phi' (0) \> = \zeta^2 (1-2\zeta^2 g_A ) \log(z\zbar/a^2)
- g^2 \zeta^8 \log^2 (z\zbar/a^2 )
\end{equation}

The $\beta{\rm eta}$-function $\beta_A (g)$ to all
orders in $g$ can be computed from (\ref{eIxxxx}) along with the 
renormalization group equation 
\begin{equation} 
\label{eIxxxxi}
\( \frac{\d}{\d \log a}  + \beta_A (g) \frac{\d}{\d g_A} \) 
\< \phi (r) \phi (0) \> = 0
\end{equation} 
One finds 
\begin{equation}
\label{eIxxxxii} 
\beta_A (g) =   2 g^2 \zeta^4 =   \frac{ 2g^2}{(1+2g)^2}, 
~~~~~\beta_g = 0  
\end{equation}
Note that this agrees with the lowest order result (\ref{eIxi}).  

We consider now correlation functions of exponentials of $\phi$
and $\phi'$.   The `free-field property' eq. (\ref{eIxxix}) 
implies that correlation functions of exponentials can be computed
using Wick's theorem in the free theory $S_\chi$.    
One has
\begin{eqnarray}
\label{expon}
\< \prod_i e^{i \alpha_i \phi (z_i, \zbar_i )} \> &=&
\prod_{i<j} \exp \( - \alpha_i \alpha_j \< \phi (z_i, \zbar_i) \phi (z_j
, \zbar_j ) \> \) 
\\
&=& \prod_{i<j} |z_i - z_j |^{2\alpha_i \alpha_j \zeta^2 (1+2\zeta^2 g_A)}
\exp \( {\alpha_i \alpha_j g^2 \zeta^8 \log^2 |z_i - z_j|^2  } \)  
\end{eqnarray} 
{}From this one deduces how the disorder affects the dimension of
the exponential  fields: 
\begin{equation}
\label{dim}
{\rm dim} \( e^{i\alpha \phi} \) =  \alpha^2 \zeta^2 
(1+2\zeta^2 g_A ) 
,~~~~~~
{\rm dim} \( e^{i\alpha \phi'} \) =  -\alpha^2 \zeta^2
(1-2\zeta^2 g_A )
\end{equation}

\subsection{ Random Field XY Model: $N$-Species }

We now extend some of the  results of the last section  to $N$-species.  
The effective action eq. (\ref{eq11N}) 
has a $gl(N|N)$ Lie superalgebra symmetry.  To see
this, define the left-moving currents 
\begin{eqnarray}
J_{ab} &=&\psi_{Lb} \psi^\dagger_{La}
  , ~~~~~~~J'_{ab} = \beta_{Lb}   \beta^\dagger_{La} 
\\ \nonumber
G_{ab} &=& \beta^\dagger_{Lb}  \psi_{La},  
~~~~~~~ G^\dagger_{ab} =  \beta_{Lb} \psi^\dagger_{La}  
\label{n4}
\end{eqnarray}
and similarly for the right-moving currents 
$\bar{J}_{ab} = \psi_{Rb}  \psi_{Ra}^\dagger$, etc.  
In the conformal field theory with $g=g_A = 0$, one has a 
$gl(N|N)$ current algebra with the operator product expansions
\begin{eqnarray}
\label{glNNkOPE}
J_{ab} (z)  J_{cd} (w) &\sim& \frac{k \ \delta_{ad}\delta_{bc}}{(z-w)^2} 
+ \inv{z-w} \( \delta_{ad} J_{bc} (w) - \delta_{bc} J_{ad} (w) \) 
\\
J'_{ab} (z) J'_{cd} (w) &\sim& 
-\frac{k \ \delta_{ad}\delta_{bc}}{(z-w)^2} 
- \inv{z-w} \( \delta_{ad} J'_{bc} (w) - \delta_{bc} J'_{ad} (w) \) 
\\
J_{ab} (z) G_{cd} (w) &\sim& \frac{\delta_{ac}}{z-w} G_{bd} (w) 
, ~~~~~~
J'_{ab} (z) G_{cd} (w) \sim  \frac{\delta_{bd}}{z-w} G_{ca} (w) 
\\
J_{ab} (z) G^\dagger_{cd} (w) &\sim& - \frac{\delta_{bc}}{z-w} G^\dagger_{ad}
 (w) 
, ~~~~~~
J'_{ab} (z) G^\dagger_{cd} (w) \sim  - \frac{\delta_{ad}}{z-w} G^\dagger_{cb}
 (w) 
\\
G_{ab} (z) G^\dagger_{cd} (w) &\sim& - \frac{k \ \delta_{ac} \delta_{bd} }
{(z-w)^2}  + \inv{z-w} \( \delta_{ac} J'_{bd} (w) - \delta_{bd} J_{ca} (w) 
\) 
\label{n5}
\end{eqnarray} 
where the level of the current algebra  is $k=1$.

The perturbing fields are current-current perturbations which preserve
the global diagonal (left-right) $gl(N|N)$ symmetry:
\begin{eqnarray}
\CO_m &=& \sum_{a,b} \( J_{ba} \bar{J}_{ab} - J'_{ba} \bar{J}'_{ab} 
+ G_{ab} \bar{G}^\dagger_{ab} - G^\dagger_{ba} \bar{G}_{ba} \) 
\\ \nonumber
\CO_A &=& - \sum_{a,b} \( J'_{aa} - J_{aa} \) \( \bar{J}'_{bb} - 
\bar{J}_{bb} \) 
\label{n6} 
\end{eqnarray} 

The structure of the perturbative expansion in $g, g_A$ about
the conformal $gl(N|N)$ current algebra closely parallels the 
1-species case.  In particular, using the OPE's (\ref{n5}), one
can verify that the  OPE's  
(\ref{eIx}) still hold.  In particular, the $N$ dependence cancels.    
Since the ultraviolet divergences arise from the singular terms 
in the OPE eq. (\ref{eIx}),   
this implies that the $\beta{\rm eta}$-functions 
are independent of $N$ and are the same as in eq.  
(\ref{eIxi}).   Thus the N-species theory
has the same $\beta$eta function
as the 1-species theory.  

Since the perturbative expansion can be derived from 
(\ref{eIx}), we expect that correlation functions in the 
$gl(N|N)$ theory have the same structure as we found for 
$gl(1|1)$.   However, since the functional integral manipulations
we used to solve the bosonized $gl(1|1)$ theory 
are not straightforwardly extended to the $gl(N|N)$ case, here
we obtain expressions for the correlation functions by using
properties of the current algebra.  This can be viewed as an alternative
method to the functional integral methods described in section IV A,
and in fact explain the results we obtained there in a more covariant
fashion.  

\def\Jb{\bar{J}}
\def\Ct{\tilde{C}}

Let $J_\alpha$ denote  an arbitrary left current, 
$J_\alpha \in \{ J_{ab}, J'_{ab} , G_{ab} , G^\dagger_{ab} \} $. 
The perturbing operators $\CO_m , \CO_A$ are $gl(N|N)$ invariant 
and can be expressed as 
\begin{equation}
\label{n7}
\CO_m = C^{\alpha \beta} J_\alpha \Jb_\beta , ~~~~~
\CO_A = - \Ct^{\alpha\beta} J_\alpha \Jb_\beta 
\end{equation}
where $C^{\alpha\beta}$ and $\Ct^{\alpha\beta}$ define two independent
Casimirs.  In the abelian sub-vector space spanned by 
$(J_{11}, J_{22}, ..., J_{NN}, J'_{11}, ..., J'_{NN} )$, one has
\begin{equation}
\label{n8}
\{ C^{\alpha \beta} \} = \left( \matrix{ 1_N & 0\cr 0& - 1_N \cr}\right) , 
~~~~~
\{ \Ct^{\alpha\beta} \} = \left( \matrix{ ~\{ 1 \} & - \{ 1 \} \cr
- \{ 1 \} & ~\{ 1 \} \cr }\right) 
\end{equation}
where $1_N$ is the $N\times N$ identity matrix, and $\{ 1 \}$ is the 
$N\times N$ matrix with $1$'s in every entry. 
Since $C$ is invertible (whereas $\Ct$ is not), $C$ defines a metric
to be utilized to raise and lower indices.  We will need 
$\Ct_{\alpha\beta} = C_{\alpha \alpha'} C_{\beta \beta'} 
\Ct^{\alpha'\beta'} $, where $C_{\alpha\beta} = C^{\alpha\beta}$.  
In the subspace defined above,
\begin{equation}
\label{n9}
\{ \Ct_{\alpha\beta} \} = \left( \matrix{ \{ 1 \} &  \{ 1 \} \cr
 \{ 1 \} & \{ 1 \} \cr }\right) 
\end{equation}
The OPE's (\ref{n5}) can be expressed as 
\begin{equation}
\label{n10}
J_\alpha (z) J_\beta (w) \sim \frac{C_{\alpha\beta}}{(z-w)^2} 
+ \inv{z-w}  f_{\alpha\beta}^\gamma ~ J_\gamma (w)
\end{equation}
where $f_{\alpha\beta}^\gamma$ are structure constants.  

The $U(1)$ currents can be bosonized as in (\ref{eIxii}, \ref{eIxiv}):
\begin{equation}
\label{n11}
J_{aa} = i\d_z \phi_a , ~~~~~\Jb_{aa}  = -i \d_\zbar \phi_a  
, ~~~~~ J'_{aa} = i \d_z \phi'_a , ~~~~~ \Jb'_{aa} = -i\d_\zbar \phi'_a 
\end{equation}
By bosonizing the fields as in (\ref{eIxiii}, \ref{eIxv} ), one
can verify that the shift property (\ref{eIxx}) continues to hold:
\begin{equation}
\label{n12}
S_{\rm eff} \[ \phi_a , \phi'_a , g_A \] = 
S_{\rm eff} \[\phi_a - \zeta^2 g_A \Sigma , \phi'_a - \zeta^2 g_A \Sigma , 
g_A = 0\]
\end{equation}
where 
\begin{equation} 
\label{n13}
\Sigma \equiv \sum_a \( \phi_a - \phi'_a \)
\end{equation}
and $\zeta$ is defined in (\ref{eIxix}).  This implies 
\begin{equation}
\label{n14}
\< \CO (\phi_a , \phi'_a ) \>_{g_A} 
= \< \CO ( \phi_a + \zeta^2 g_A \Sigma, \phi'_a + \zeta^2 g_A \Sigma )
\>_{g_A = 0} 
\end{equation}
where $\CO$ is any operator.  

The field $\Sigma$ has very simple correlation functions because it
doesn't appear in any exponential interactions.  Collecting the kinetic
terms for the scalar fields coming from the conformal field theory
and from the terms $g(J_{aa} \Jb_{aa} - J'_{aa} \Jb'_{aa})$ 
\begin{equation}
\label{n15}
S_{\rm kinetic} =  {1\over \zeta^2 } \int \frac{d^2x}{4\pi} 
\sum_a \d_z \phi^+_a \d_\zbar \phi^-_a 
\end{equation}
where
\begin{equation}
\label{n15b}
\phi^\pm_a = \phi_a \pm \phi'_a 
\end{equation}
Since $\Sigma$ doesn't couple to any of the remaining interaction
terms, one has the exact 2-point function
\begin{equation}
\label{n16}
\< \phi^+_a (z,\zbar) \Sigma (0) \> = - 2 \zeta^2 \log z\zbar 
\end{equation}

The above properties, together with current conservation provide strong
constraints on the 2-point functions of the currents.  Euclidean
rotational invariance, and the fact that $J_\alpha , \Jb_\alpha$ are
dimension $1$ currents constrains the $\langle J \Jb \rangle$
2-point function to be of the form:
\begin{equation}
\label{n17}
\< J_\alpha (z ,\zbar ) \Jb_\beta (0) \>_{g_A = 0} 
= \inv{z\zbar}  \( C_{\alpha\beta} F(g, z\zbar \mu^2) 
+ \Ct_{\alpha\beta} \tilde{F} (g, z\zbar \mu^2 ) \) 
\end{equation}
where we have used the unbroken $gl(N|N)$ symmetry.  
$F, \tilde{F}$ are scaling functions of $g$ and the dimensionless
combination $z\zbar \mu^2$, where $\mu$ is the energy scale that
enters through the renormalization group.  

The field $\Sigma$ has the covariant description:
\begin{equation}
\label{n17b}
i\d_z \Sigma = \Ct_\alpha^\beta J_\beta, ~~~~~
-i \d_\zb \Sigma = \Ct_\alpha^\beta \Jb_\beta, ~~~~~ \forall~ \alpha
\end{equation}
{}From (\ref{n16}) we see that away from $z\zb = 0$, one must have 
$\Ct_{\alpha'}^\alpha \langle J_\alpha \Jb_\beta \rangle = 0$.  
Since $\Ct_{\alpha'}^\alpha C_{\alpha\beta} \neq 0$, 
and $\Ct_{\alpha'}^\alpha \Ct_{\alpha\beta} = 0$, one
deduces that $F=0$.  

The property (\ref{n16}), together with (\ref{n14}), leads to 
$g_A$ independence (up to $\delta$-functions) of the correlation
function (\ref{n17}).  Since the dependence on the scale 
$\mu$ must be compatible with the $\beta{\rm eta}$-function for $g_A$, the
$g_A$ independence implies that $\tilde{F}$ is only a function 
of the dimensionless coupling $g$.  Thus
\begin{equation}
\label{n19}
\< J_\alpha (z,\zbar) \Jb_\beta (0) \>_{g_A = 0} 
= \inv{z\zbar} \Ct_{\alpha\beta} \tilde{F} (g) 
\end{equation}

Current conservation, $\d_\zbar J_\alpha + \d_z \Jb_\alpha = 0$, 
requires 
\begin{equation}
\label{n20}
\d_\zbar \< J_\alpha (z, \zbar) J_\beta (0) \> 
= - \d_z \< J_\alpha (z , \zbar ) \Jb_\beta (0) \> 
\end{equation}
Integrating this, 
\begin{equation}
\label{n21}
\< J_\alpha (z, \zbar) J_\beta (0) \>_{g_A = 0}  
= \inv{z^2} \( \zeta^2 C_{\alpha\beta} + \Ct_{\alpha\beta} \tilde{F} (g) 
\log z\zbar \) 
\end{equation}
The coefficient $\zeta^2$ of the $C_{\alpha\beta}$ term was derived from
(\ref{n16}).

To fix the one unknown function $\tilde{F} (g)$, we use current
algebra equations of motion.  The equations of motion to first
order in $g$ can be obtained in conformal perturbation theory
\begin{equation}
\label{n22}
\d_\zbar J_\alpha (z,\zbar) = g \oint_z \frac{dw}{2\pi i} 
J_\alpha (w, \zbar) \CO_m (z, \zbar ) 
\end{equation}
Taking into account the $\zeta$-rescaling $\phi \to \phi/\zeta$ that 
gives the kinetic term the standard normalization, and using the 
OPE's (\ref{n10}), one obtains 
\begin{equation}
\label{n23}
\d_\zbar J_\alpha = g \zeta^2 f_\alpha^{\beta\gamma} J_\beta \Jb_\gamma
= - \d_z \Jb_\alpha 
\end{equation}
Using this in the correlation function  
 $\langle \d_\zbar J_\alpha \d_z \Jb_\beta \rangle$, along with 
(\ref{n17}), 
\begin{equation}
\label{n24}
\inv{z^2 \zbar^2} \Ct_{\alpha\beta} \tilde{F} 
= g^2 \zeta^4 f_\alpha^{\beta'\gamma} f_\beta^{\sigma\gamma'} 
\< (J_{\beta'} \Jb_\gamma ) (z,\zb) ( J_\sigma \Jb_{\gamma'} ) (0)
\>_{g_A = 0} 
\end{equation}
The non-zero contribution to $\tilde{F}$ comes from 
\begin{equation}
\label{n25} 
\< (J_{\beta'} \Jb_\gamma ) (z,\zb) ( J_\sigma \Jb_{\gamma'} ) (0)
\>_{g_A = 0}  \sim \frac{\zeta^4}{z^2 \zbar^2} C_{\beta' \sigma}
C_{\gamma\gamma'} 
\end{equation}
where we have used (\ref{n21}).  Now, in terms of the structure constants,
the OPE  $\CO_m (z,\zb) \CO_m (0) \sim -2 \CO_A /z\zbar $ implies 
\begin{equation}
\label{n27}
2 \Ct_{\alpha\beta} = f_\alpha^{\beta'\gamma} f_\beta^{\sigma\gamma'}
C_{\beta'\sigma} C_{\gamma\gamma'} 
\end{equation}
The last three equations thus give 
\begin{equation}
\label{n26}
\tilde{F} (g) = 2 g^2 \zeta^8
\end{equation}

Let us summarize the above results by integrating the current correlations
to find the 2-point functions of the scalar fields:
\begin{eqnarray}
\label{n29} 
\< \phi^+_a (z,\zbar) \phi^+_b (0) \>_{g_A = 0} 
&\sim& -4 g^2 \zeta^8 \log^2 z\zbar, ~~~~~ \forall ~ a,b 
\\
\< \phi^+_a (z,\zbar) \phi^-_b (0) \>_{g_A = 0} 
&\sim& -2  \zeta^2  \delta_{ab} \log z\zbar 
\\
\< \phi^-_a (z,\zbar) \phi^-_b (0) \>_{g_A = 0} 
&\sim&  0 , ~~~~~\forall ~ a,b
\end{eqnarray}
Finally, using the shift property (\ref{n14}), 
\begin{equation}
\label{n30}
\< \phi_a (z,\zbar) \phi_b (0) \> = 
-\zeta^2 \( \delta_{ab} + 2 \zeta^2 g_A \) \log z\zbar 
- g^2 \zeta^8 \log^2 z\zbar
\end{equation}
\begin{equation}
\label{n30bos}
\< {\phi'}_a (z,\zbar) {\phi'}_b (0) \> = 
\zeta^2 \( \delta_{ab} - 2 \zeta^2 g_A \) \log z\zbar 
- g^2 \zeta^8 \log^2 z\zbar
\end{equation}

\subsection{  Nearly Conformal Structure}

We have seen in the previous section that the N-species theory
(describing $N$ copies of the random field XY model)
is nearly conformal. This can be seen more clearly by
separating out the non-conformal pieces (coupling holomorphic
and anti-holomorphic coordinate dependences) by  rewriting
Eq.'s(\ref{n19},\ref{n21}) as follows.
Instead of the basis of currents as in
Eq.(\ref{n4}),
we extract   $U(1)\times U(1)$ 
from the bosonic $U(N)\times U(N)$ subgroup.
Explicitly, consider traceless generator matrices  $T_A$ of $su(N)$, 
($A=1,...,N^2-1$) 
$$
J_A\equiv (T_A)_{ab} J_{ab}, \qquad \qquad 
{J'}_A\equiv (T_A)_{ab} {J'}_{ab} 
$$
and
$$ 
J\equiv \sum_{a} J_{aa},
\qquad
{J'}\equiv \sum_{a} {J'}_{aa}
$$
and similar for the fermionic currents
(which we denote by $J_A$, $A=2N^2+1, ..., 4N^2$).
Forming the combinations
$$
J_-\equiv J -{J'},
\qquad
J_+\equiv J +{J'},
$$
and noting that all non-conformal terms  in the current-current
 correlators are multiplied by the 
invariant ${\tilde C}_{AB}$ whose only non-vanishing matrix elements
are  $A,B=\pm$ in this basis,
we see that the two point correlation functions of all the remaining
currents are
$$
<J_{A} (z,{\bar z}) J_B(0)>
= {\zeta^2 C_{AB}\over z^2},
\quad
<J_{A} (z,{\bar z}) J_{\pm}(0)> \equiv 0
\qquad \qquad ( A,B \not = \pm)
$$
Note that we see from Eq.(\ref{n25}) 
that these correlators are invariant under independent action of the symmetry
group on the left- and right- chiral components of the
currents.

The ability to separate off the $J_{\pm}$ currents (and the
corresponding fields, $\phi_{\pm}$) can be seen from the
hamiltonian,  expressed as
$$
H=H_0 + H_I
$$
where $H_0$ is the non-interacting $gl(N|N)$-Sugawara Hamiltonian,
and $H_I$ is the current-current interaction
Recalling that $C_{AB}=C^{AB}$ there are no terms in the
hamiltonian $H$ which couple the generators $J_{\pm}, {\bar J}_{\pm}$
to the remaining ones, which allows to
consistently set
$J_{\pm}= {\bar J}_{\pm}=0$, yielding a
scale invariant theory.

Setting $J_\pm = \bar{J}_\pm = 0$ effectively sets $g_A = 0$.  Since
the non-zero $\beta$eta-function is $\beta_{g_A} (g)$, what is left
is a conformal field theory since $\beta_g = 0$.  In terms of groups,
setting $J_\pm = 0$ is equivalent to dividing by the $U(1)\otimes U(1)$
subgroup:  $GL(N|N)/ U(1) \otimes U(1) = PSL(N|N)$.  Since the two
$U(1)$ bosons give $c=2$, and the total central charge for $GL(N|N)$ is
zero, what is left is a $c=-2$ conformal field theory.
The conformal $PSL(N|N)$ sigma model was recently
 studied in \cite{Bershadsky}\cite{Witten} 
(see also section VI  below).

In the $N=1$-species case the emergence of conformal 
symmetry is easily understood: Bosonizing
the non-interacting theory, there are only two
generators $G_+=i\partial \chi$, $G_-=i\partial \chi^\dagger$
left once the currents $J_{\pm}$ and thus the fields
$\phi_{\pm}$ have been eliminated.
The $PSL(1|1)$ sigma model is  a  free theory.

One proceeds similarly for the $N>1$ cases. 
The $N=2$ species case for example
describes the second moments of observables in
the random XY model. An interesting observable is the
`Edwards-Anderson order parameter',
which is a bilinear of one of the `conformal' currents:
$$ 
q_{12}\equiv 
(\psi^{\dagger}_{L1}\psi_{R1})
( \psi_{L2} \psi^{\dagger}_{R2})
=
J_{12} {\bar J}_{21},
$$
We see from Eq.(\ref{n25})
that the two-point  of this field decays 
algebraically,
$$
<q_{12}(z,{\bar z}) q_{21}(0)>
\propto
{\zeta^4\over
z^2 {\bar z}^2 }
$$

It was recently proposed by Zirnbauer that the $PSL(N|N)$ sigma models
describe the integer Quantum Hall transition\cite{Zirnbauer}.
The kind of disorder considered in the present paper, which,  as we have explained,
leads to the $PSL(N|N)$ sigma models, is not generally believed to
correspond to kind of disorder needed for the Quantum Hall effect,
at least not in an obvious way. Nevertheless this is an interesting proposal
that needs further investigation.

\subsection{  Generalized Random XY Model (Broken Time-Reversal Symmetry)}

In this  subsection we discuss a more general random   XY model 
(at the free fermion point).
This model may be defined in terms of an associated 
$2$-component Dirac  hamiltonian of type $H_2$, where now  all random potentials
have
{\it both} real  ($A_x, A_y, V, M$)  
and imaginary (${A'}_x, {A'}_y,{V'},  {M'}$)  parts.
As discussed in Appendix B, in this case  the  corresponding $4$-component
hamiltonian $H_4$ lacks in general time reversal symmetry (but still  has
particle-hole symmetry), and the corresponding field theory
action possesses $GL(1|1;C)$ global SUSY\footnote{
$GL(N|N;C)$ symmetry for the  $N$th moment averages.}. 
This subsection is devoted to the
1-loop RG equations of this theory.
To summarize the result, we find that the theory  
flows off to strong coupling. The strong coupling
physics  should be captured by a  non-linear sigma model of the kind discussed
in Section  VI.

Performing the disorder average over 
independent random variables
$m=V+iM$ and $\mu=i(V'+iM')$  with variance $g_m=\overline{m m^*}$ and
$g_\mu=\overline{\mu \mu^*}$ respectively\footnote{
using the action of Eq.(\ref{appsusyactionLR}) of Appendix B},
we arrive at the following   lagrangian for the averaged theory
\begin{eqnarray}
\nonumber
{\cal {\bar L}}^g_{susy}
&=&
g_1
 \sum_{a=1,2} \( J_{aa} \bar{J}_{aa} - J'_{aa} \bar{J}'_{aa} 
+ G_{aa} \bar{G}^\dagger_{aa} - G^\dagger_{aa} \bar{G}_{aa} \) 
\\ 
\label{Lbarsusy}
&~& ~~~~~+
g_2
 \sum_{a,b=1,2}^{a\not = b} 
\( J_{ba} \bar{J}_{ab} - J'_{ba} \bar{J}'_{ab} 
+ G_{ab} \bar{G}^\dagger_{ab} - G^\dagger_{ba} \bar{G}_{ba} \) 
\end{eqnarray}
where
\begin{equation}
\label{DEFgonegtwo}
g_1\equiv g_m-g_{\mu},
\qquad
g_2\equiv g_m+g_{\mu}
\end{equation}
The additional contribution from the random vector potentials is
\begin{equation}
\label{LbarA}
{\cal {\bar L}}_{susy}^A = 
g^A_1
(-1)\sum_{a=1,2} \ 
({J'}_{aa}-J_{aa}) 
({{\bar J}'}_{aa}-{\bar J}_{aa}) 
+
g^A_2
(-1)\sum_{a=1,2}^{a\not = b} \ 
({J'}_{aa}-J_{aa}) 
({{\bar J}'}_{bb}-{\bar J}_{bb}) 
\end{equation}
where
\begin{equation}
\label{DEFgAgAprime}
g^A_1= g_A - g_{A'},
\qquad
g^A_2= g_A + g_{A'}
\end{equation}
[$g_A$ and $g_{A'}$ are the variances of the
imaginary and the real vector potential, respectively.]

We recover the time-reversal symmetric case, 
Eq.(\ref{n6}),
 by letting $g_\mu=g_{A'}=0$.
We note that in the special case where only the potentials
$A_x, A_y, {V'}, {M'}$  
are non-vanishing,
which corresponds to $g_m=0$, $ g_\mu>0$,
 we recover time reversal symmetry [Eq. (\ref{Tprime})
of Appendix B]. 
The RG equations for this case are identical to those
of the case $g_m>0, g_\mu=0$, studied in the the previous section,
upon letting $g_m \to - g_\mu$.
This can be seen by
replacing in  Eq.(\ref{DEFbetagamma}, \ref{DEFpsiLR})
the left moving fields
$\psi_{L1}$  by 
$i\psi_{L1}$,  $\psi_{L2}$ by 
$-i\psi_{L2}$ 
( and
$\psi^\dagger_{L1}$ by $-i\psi^\dagger_{L1}$, and
 $\psi^\dagger_{L2}$ by $i\psi^\dagger_{L2}$)
and similarly for the  $\beta$-fields. All 
right moving fields remain unchanged.
With these redefinitions the coupling constants
in Eq.(\ref{Lbarsusy}) become 
$g_1=g_2=-g_\mu$.
On the other hand, as these redefinitions
do not change the kinetic term in
Eq.(\ref{appsusyactionLR}) of Appendix B,
the OPE's of these fields remain unchanged.
This implies that the  OPE's of the $gl(2|2)$ currents in Eq.
(\ref{n5}) remain also unchanged, and
the entire analysis of
sections IV A, B  remains unchanged upon replacing
$g_m$ by $-g_\mu$.

Let us now return to the  {\it most general case}.
The 1-loop RG equations for the coupling constants of the action above
are derived in Appendix C
( Eq.'s (\ref{dgonedgtwo}), (\ref{dgAdgAprime}) 
with the result\footnote{
In terms of the notations of the Appendix, 
$D\equiv  ( d^2_{22} - d^2_{21} )$}:
$$
{d g_1\over dl} =0,
\qquad
{d  g^A
\over d l}
=
2 [(g_m)^2 + (g_\mu)^2]
$$

\begin{equation}
\label{RGnotimerev}
{d g_2\over dl}
=
D  g_{A'}
  g_2,
\qquad 
{d  g^{A'}
\over d l}
=
4  g_m
 g_\mu 
\end{equation}
The coupling $g_2$ flows according to
$$
{ d^2 \log (g_2)\over d l^2}
 =
D
 { d  g_{A'} \over dl}
=
D
 [(g_2)^2-(g_1)^2]
$$
which means that $g_2$ flows to large values.
( Note from Eq. (\ref{DEFgonegtwo})
that the expression
$ [(g_2)^2-(g_1)^2]$ is always non-negative.)
The strong coupling physics should be described  via the sigma model
technology of section VI.

\section{Density of States of  the  Hatsugai et al. 
Delocalization Transition from Current Algebras}

In this section we apply some of the above results to the problem
of localization of electrons hopping with {\it real}
amplitudes  on a square lattice
with flux $\pi$ per plaquette \cite{hatsugai}.
In the continuum limit the hamiltonian of this system is the
4-component (hermitian)   Dirac hamiltonian $H_4$, 
\begin{equation}
\label{h4}
H_4
=\pmatrix{ 0 & H_2 \cr
 H_2^\dagger & 0  \cr}
\end{equation}
where $H_2$ is defined in eq. (\ref{eq1}).
The reality of lattice hopping amplitudes implies
that the hamiltonian $H_4$  must have 
time-reversal symmetry, which is  manifest with this form of $H_2$
(see Eq.(\ref{Treversal})).
The corresponding SUSY field theory action  (Eq.(\ref{appsusyactionLR})
of Appendix B with $\mu \equiv  0$)
can be seen to possess $GL(2|2;R)$ symmetry.
(This model has also  been studied numerically in \cite{Morita}.)
In this subsection we compute the density of states of the
hamiltonian $H_4$.
In particular, we  will be  interested in the eigenstates of the Schr\"odinger
equation $H_4 {{\Psi}} = E {{\Psi}}$, where ${\Psi}$
is a 4-component wave function.  The single particle
Green functions are defined by the functional integral
$\int D{\bf{\Psi}}^* D{\bf{\Psi}} ~ \exp ( - S_0^f ) $,
where
\begin{equation}
\label{eq3}
S_0^f = \int \dtx {\bf{\Psi}}^\dagger (x) i \( H_4 - \CE \) {\bf{\Psi}} (x)
\end{equation}
and $\CE = E + i\ep$.
For $\ep = 0^+$, the functional integral defines the
retarded Green function:
\begin{equation}
\label{eq4}
{G_R (x,x';E)}_{ab} = \lim_{\ep \to 0^+} \langle x, a|
\inv{H_4 - (E + i\ep)} |x', b \rangle
= \lim_{\ep \to 0^+}  i \langle 
{\bf{\Psi}}_a 
 (x) 
{\bf{\Psi}}^\dagger_b
(x') \rangle
\end{equation}
Making the identifications\footnote{The superscript ${}^t$ denotes
the transposed vector.} (Appendix B)
\begin{equation}
\label{eq5two}
{\bf{\Psi}} = ( \psi_{L2}^\dagger , \psi_{R2}^\dagger,
 \psi_{L1} , \psi_{R1} )^t,
~~~~~~
{\bf{\Psi}}^\dagger = ( \psi_{R1}^\dagger, \psi_{L1}^\dagger ,
\psi_{R2},  \psi_{L2})
\end{equation}
 the action takes the form of an $\CE$ perturbation of the 
$N=2$ species model:
\begin{eqnarray}
S  &=& 
\int \dtx {\bf{\Psi}}^\dagger (x) i \( H_4 - \CE \) {\bf{\Psi}} (x)
\\ \nonumber
&=& 
\int \dtx ~   \sum_{a=1}^2 \(
\psi_{La}^\dagger  \d_\zbar \psi_{La} +
\psi_{Ra}^\dagger \d_z \psi_{Ra}   
+ m(x) \psi_{Ra}^\dagger \psi_{La} + {m^*} (x) \psi_{La}^\dagger \psi_{Ra} \)
\\ \nonumber
&~& ~~~~
+  A_\zbar \( \psi_{L1}^\dagger \psi_{L1} + \psi_{L2}^\dagger \psi_{L2} \)
+  A_z \( \psi_{R1}^\dagger \psi_{R1} + \psi_{R2}^\dagger \psi_{R2} \)
  ~~-i \CE \int  \( \psi_{L1}^\dagger \psi_{R2}^\dagger - \psi_{L2}^\dagger
\psi_{R1}^\dagger + h.c. \)
\label{twospec}
\end{eqnarray}

For simplicity, we can set the physical $A$-couplings 
to zero from the beginning
and then let the $g_A$ coupling be 
  generated dynamically under
renormalization.  
Introducing ghosts and integrating over disorder leads to the
effective action:
\begin{equation}
S = S_{\rm eff}^{(N=2)} ~  -i \CE \int \dtx ~ \CO_E 
\end{equation}
where 
\begin{equation}
\label{coe}
\CO_E = \psi_{L1}^\dagger \psi^\dagger_{R2} - 
\psi_{L2}^\dagger \psi^\dagger_{R1}
 + \beta_{L1}^\dagger \beta^\dagger_{R2} + 
\beta_{L2}^\dagger \beta^\dagger_{R1} + {\rm `h.c.'}
\ \  {\rm terms}
\end{equation}
(see Appendix B).

Of interest is the density of states (DOS)
\begin{equation}
\label{Div}
\rho(E) = \inv{\CV} {\rm Tr} \delta (H_4-E) = \inv{\CV\pi}
\lim_{\ep \to 0^+} {\rm Im} {\rm Tr} \( \inv{H_4-E-i\ep} \)
\end{equation}
where $\CV$ is the two-dimensional volume.  The averaged density
of states can be expressed in
terms of the averaged retarded Green function
\begin{equation}
\label{Dv}
\overline{\rho(E)} = \inv{\pi} \lim_{\ep \to 0^+} {\rm Im} \   tr \ 
\overline{ G_R (x,x;E)}
\end{equation}
where $tr$ denotes the trace over matrix indices 
of the Green's function of Eq.(\ref{eq4}).

Thus the density of states is related to the one-point function of
the, say, fermionic part of the  operator $\CO_E$ which couples to $\CE $:
\begin{equation}
\label{Dix}
\bar{\rho (E)} = \inv{\pi} \lim_{\ep \to 0^+} ~ 
Re \langle i (  \psi^\dagger_{L1}\psi^\dagger_{R2}
-\psi^\dagger_{L2}\psi^\dagger_{R1}
 )
\rangle_{\CE} 
\end{equation}
where the one-point function on the right hand side is computed using
$S_{\rm eff}$ with the $\CE$ term.

Generally, the $\CE$-term spoils the complete solvability of the theory. 
Our strategy will be to use our solution of the $\CE = 0 $ theory
in conjunction with the renormalization group to learn something
about the density of states. 
Let us view $S_\CE = -i\CE \int \dtx \CO_E $ as a perturbation of
the nearly conformal field theory.   The anomalous
scaling dimensions of fields are controlled by $S_{\rm eff}$ when
$\CE = 0$.

We need the anomalous dimension $\Gamma$ of
the field
$\CO_E$ which contains
$
: \psi_{L1}^\dagger \psi_{R2}^\dagger :
 = : \exp ( -i\varphi_{L1} 
+ i \varphi_{R2} ) :
= $ 
$\exp\{-i({\tilde \phi}_1 +{\tilde \phi}_2)/2 \}
\  \exp\{-i( \phi_1 -\phi_2)/2\}$.
Here, ${\tilde \phi}_a=
\varphi_{La} -\varphi_{Ra}$
denotes the dual field, whose 2-point function
is found from
Eq.(\ref{n19}), (\ref{n26}) to be
$$
<{\tilde \phi}^+_a(z,{\bar z}) {\tilde \phi}^+_b(0)>_{g_A=0}
\sim
4g^2\zeta^8 \log^2(z{\bar z}/a^2)
$$
The requirement that for arbitrary $g_A$ this function is to satisfy
Eq.(\ref{eIxxxxi})
gives
$$
<{\tilde \phi}^+_a(z,{\bar z}) {\tilde \phi}^+_b(0)>_{g_A}
\sim
8\zeta^4 g_A  \log (z{\bar z}/a^2)
+
4 g^2\zeta^8 \log^2(z{\bar z}/a^2)
$$
This together with
$$
<{\tilde \phi}^+_a(z,{\bar z}) {\tilde \phi}^-_b(0)>_{g_A}
\sim
-2\zeta^2 \delta_{ab} \log (z{\bar z}/a^2)
$$
yields the anomalous dimension\footnote{For $N \geq 2$ the free field
property that leads to exact expressions such eq. (\ref{expon}) is not
expected to hold.  However if  we assumed that
 for the purposes of computing
the anomalous dimension we  can parallel what we did for $N=1$,
we would obtain $h(g)=\zeta^2/2$.} 
of the field $: \psi_{L1}^\dagger \psi_{R2}^\dagger :$
$$
\Gamma(g_A)
={1\over 2}\zeta^2 (1-4\zeta^2 g_A) + h(g)
$$
where $h(g)$ is independent of $g_A$.
Since the action is dimensionless,
\begin{equation}
\label{Dxi}
 {\rm dim} (E) = 2- \Gamma
\end{equation}
ensuring  that $\bar{\rho (E)} dE$ has
dimension $2$, as it should in two dimensions.

The renormalization group equation for $\bar{\rho} (E, g_A)$ 
\begin{equation}
\label{Ri}
\( (2-\Gamma) E \frac{\d}{\d E} + \beta_A (g) \frac{\d}{\d g_A}
 - \Gamma(g_A) \) \bar{\rho} (E, g_A) = 0
\end{equation}
implies that 
the renormalized density of states ${\bar \rho}(l)\equiv
\bar{\rho} (E(l), g_A(l))
$
(where $e^l \geq 1$ is the rescaling factor) is related
to the unrenormalized  DOS ${\bar \rho}(0)$ by
$$
{\bar \rho}(0)/{\bar \rho}(l)=
\exp 
\{
-\int_0^l \Gamma(l') dl'
\}=
\exp 
\{
\int_0^l \ [2-\Gamma(l')] dl'
-2l
\}
$$
Here
$$
{ d E(l) \over dl}
=[2-\Gamma] E,
\qquad
{d g_A(l) \over dl}
= \beta_{g_A} 
$$
which  permits to express the $l$-dependence
in terms of the
 the renormalized coupling constant $E_R\equiv E(l)$
and its bare value $E \equiv E(0) \leq E_R$.
This yields for the DOS
\begin{equation}
\label{R11}
\bar{\rho (E)}  /\bar{\rho (E_R)}= 
{E_R \over E}
 \exp \( -2 [g_A (E_R)-g_A (E)]
 / \beta_{g_A} \) 
\end{equation}
(valid since $\beta_{g_A}(g,g_A)$ is independent of $g_A$).
{}From this we find in the limit $E/E_R\to 0$
$$
{\bar \rho}(E)/{\bar \rho}(E_R)
\sim
{E_R \over E}
 \ 
\exp \{
-{\sqrt{2}(1+2g)^2\over g}
\sqrt{\log (E_R/E)} \}
$$
This diverging (but integrable)
 density of states was not seen in the simulations
of \cite{Morita}. Rather, a power law varying with the strength
of disorder $g$ was found. Such power law behavior of the DOS
is obtained from the R.G. analysis above for intermediate
not asymptotically small energies.

\section{ Gade-Wegner Delocalization Transition 
via $GL(N|N;C)/U(N|N)$ Sigma Model}

\subsection{ $GL(1|1;C)/U(1|1)$ Sigma Model}

Localization problems are usually described
in terms of sigma models, using replicas   
\cite{Wegner}\cite{Stone}  or supersymmetry\cite{Efetov}.

In this section we study the Gade-Wegner localization problem\cite{Gade}.
A version of this model consists of electrons hopping
on a 2-dimensional square lattice with arbitrary 
{\it complex} hopping amplitudes (and flux $\pi$ per plaquette).
The continuum limit of the hamiltonian of this model
is again, as for the model discussed in section V,
 of the form of a $4$-component Dirac hamiltonian.
In contrast to the latter model, however,
the lattice hamiltonian of the present model, and thus also its continuum
limit $H_4$ lacks time reversal invariance due
to the {\it complex} hopping amplitudes.
As discussed in more detail in 
Appendix B, the associated Dirac hamiltonian $H_4$
has now  arbitrary {\it complex} random 
scalar potential ( with real and imaginary parts $V'$ and $V$)
 and  {\it complex} Dirac mass ( with real and imaginary parts $M$ and $M'$) terms,
which may be combined into two independent complex
random variables $m$ and $\mu $, with variance $g_m$
and $g_\mu$ respectively.
The explicit form of the hamiltonian is
(from Eq.(\ref{appiH4}) of Appendix B):
\begin{equation}
   \ H_4=
(-i)\pmatrix{
\epsilon & 0 &  m - \mu & \partial  \cr
0 & \epsilon & {\bar  \partial}    & m^*+\mu^* \cr
-m^*+\mu^* & \partial   & \epsilon & 0 \cr
{\bar \partial}   & -m-\mu & 0 & \epsilon\cr
}
\end{equation}
where
$$
m\equiv (V+iM),
\qquad \qquad
\mu \equiv i  (V'+iM')
$$
In general, an arbitrary complex 
vector potential 
will be generated upon RG flow
in the corresponding SUSY field theory\footnote{see
 Eq. (\ref{RGnotimerev}).}.
 We set the vector potentials to zero
initially, and let it be generated upon renormalization
group transformations later on.
In order to study this system we use the conventional 
sigma-model technique.

The lagrangian is 
\begin{equation}
\label{sig1} 
\CL = \Psi^\dagger i ( H_4 - \CE ) \Psi  + \Phi^\dagger i (H_4 - \CE ) \Phi 
\end{equation}
where  $\CE = E + i \epsilon $ ($\epsilon = 0^+$) and
\begin{equation}
\label{sig2} 
\Psi = \left( \matrix{\psi_1 \cr \psi_2 \cr \psi_3 \cr \psi_4 \cr} \right) 
, ~~~~~
\Phi = \left( \matrix{\phi_1 \cr \phi_2 \cr \phi_3 \cr \phi_4 \cr} \right) 
\end{equation} 
\def\Up{\Upsilon}
\def\Sig{{\tilde{\Upsilon}}} 
Let us define new fields 
\begin{equation}
\label{sig3} 
\Up =
 \left( \matrix{\psi_1 \cr \phi_1 \cr \psi_4 \cr  \phi_4 \cr } \right)
 =\left( \matrix{ \Up_+ \cr \Up_- \cr } \right)
, ~~~~~
\Sig = 
\left( \matrix{\psi_3 \cr \phi_3 \cr -\psi_2 \cr - \phi_2 \cr } \right)
=\left( \matrix{\Sig_+ \cr \Sig_- \cr } \right)
\end{equation}
\begin{equation}
\Up^\dagger =
 \left( \matrix{\psi_1^*, &   \phi_1^*, &   
\psi_4^*,  &    \phi_4^* \cr } \right)
 =\left( \matrix{ \Up_+^\dagger, & \Up_-^\dagger \cr } \right)
 , ~~~~~
\end{equation}
\begin{equation}
\label{sig3dagger} 
\Sig^\dagger = 
\left( \matrix{\psi_3^*,  &  \phi_3^*,  &  
-\psi_2^*,  &  - \phi_2^* \cr } \right)
=\left( \matrix{ \Sig_+^\dagger, & \Sig_-^\dagger \cr } \right)
\end{equation}
Using  Eq.(\ref{appsusyactionUpsilon}) of 
the Appendix
the lagrangian can be expressed as 
\begin{equation}
\label{sig4} 
\CL =  \Sig^\dagger D \Sig - \Up^\dagger D \Up + 
m_i(x) \Up_i^\dagger \Sig_i 
- m_i^* (x) \Sig_i^\dagger \Up_i  
- i \CE \( \Sig^\dagger \Sig + \Up^\dagger \Up \) 
\end{equation} 
\begin{equation}
\label{mi}
m_{\pm}
\equiv
(m\pm\mu)
\end{equation}
where a sum over the repeated index $i=\pm 1$ is implied and
\begin{equation} 
\label{sig5}
D = (\sigma_+ \otimes 1 ) \d_z + (\sigma_- \otimes 1) \d_\zbar 
= \left( \matrix{ 0&0&\d_z &0\cr 0&0&0&\d_z \cr 
\d_\zbar & 0 &0 &0 \cr 0&\d_\zbar &0&0\cr } \right) 
\end{equation}

\def\G{{\bf \CG}}
\def\Gt{{ \bf {\tilde{\CG}} }}

When $\CE = 0$, the lagrangian has the following symmetry:
\begin{equation}
\label{sig6} 
\Up \to  \G   \Up = \left( \matrix{ (G^{-1})^\dagger & 0 \cr 0 & 
G \cr }
\right) ~ \Up 
, ~~~~~~~~
\Sig \to \Gt  \Sig =
  \left( \matrix{ G&0\cr 0& (G^{-1})^\dagger \cr } \right) \Sig
\end{equation}
where $G$ is a $2\times 2$ supermatrix with bosonic (fermionic) 
diagonal (off-diagonal) elements.   That this is a symmetry
follows from  $ \G^\dagger (\sigma_\pm \otimes 1) \G  =
\Gt^\dagger (\sigma_\pm \otimes 1 ) 
\Gt  = \sigma_\pm \otimes 1 $
and $\Gt^\dagger \G  = 1$.  Since $G$ is only required to be 
invertible, $G$ is an element of $GL(1|1; C)$, i.e. the complexification 
of $GL(1|1)$.  In the presence of the $\CE$ term, the above transformation
continues to be  a symmetry
only if $\G^\dagger \G  = \Gt^\dagger \Gt = 1$, which implies   
$G^\dagger G = 1$.  Thus the $\CE$ term breaks the symmetry 
to $U(1|1)$.   

\def\Qt{\tilde{Q}}
\def\Mt{\tilde{M}} 

Without loss of generality the random variables $(m_+,m_+^*)$
and $(m_-,m_-^*)$
may be taken to be statistically 
independent\footnote{For general values of $g_m$ and $g_\mu$,
the variables $m_+$ and $m_-$ will be correlated and will have different
variances. Changing the ratio $g_m/g_\mu$
will however  not affect
the universality class of the sigma model.
Therefore we may chose this ratio to be unity.} 
with the distribution 
(\ref{eIii}) each, and the same variance $g$.
Upon disorder averaging
 one then  obtains the term in the effective lagrangian: 
\begin{equation}
\label{sig0} 
\CL_{eff} = g  \sum_{i=\pm}
 (\Up_i^\dagger \Sig_i) 
(\Sig_i^\dagger \Up_i ) 
\end{equation}

Introduce a grade $[a]$ for the vector index $a$, where $[1] = [3] = 1$ 
(fermions),
$[2] = [4] = 0$ (bosons), such that 
\begin{equation}
\label{sig7}
\Sig_a \Up_b = (-1)^{[a][b]} \Up_b \Sig_a 
\end{equation}
Then using the fact that $(-1)^{[b]^2 + 2[a][b] } = (-1)^{[b]} $, 
one finds 
\begin{equation}
\label{sig8} 
 (\Up^*_{i p}  \Sig_{i p}  )(\Sig^*_{i q}  \Up_{i q}  ) 
= (-1)^{[q]} (\Up_{i q}  \Up^*_{i p}  ) 
(\Sig_{i p}  \Sig^*_{i q}  ) = {\rm Str } M_i \Mt_i 
\end{equation}
where ${\rm Str}$ denotes the supertrace and  $M_i, \Mt_i$ are the $2\times 2$
matrices of bilinears:
\begin{equation}
\label{sig9} 
M_i = \Up_i \Up_i^\dagger, ~~ \Mt_i = \Sig_i \Sig_i^\dagger;
~~~~~~~~{(M_i)}_{pq} = \Up_{ip}  \Up_{i q}^* 
, ~~ {(\Mt_i)}_{pq} = \Sig_{i p} \Sig^*_{i q} , ~~~~ p,q =1,2
\end{equation}

Let us introduce two $4\times 4$ block diagonal  
supermatrices of fields $Q, \Qt$
$$
Q
\equiv
\pmatrix { Q_+ & 0 \cr
            0 & Q_-\cr},
\qquad
\qquad
{\tilde Q}
\equiv
\pmatrix { {\tilde Q_+} & 0 \cr
              0 & {\tilde Q_-}\cr}
$$
and define
$$
M\equiv \pmatrix{M_+ & 0 \cr
0 & M_-\cr},
\qquad \qquad
{\tilde M}\equiv
\pmatrix{
{\tilde M}_+ & 0 \cr
0 & {\tilde M}_-\cr}
$$

Consider the following lagrangian:
\begin{equation}
\label{sig10}
\CL = {\rm Str} \( (D-g\Qt - i\CE ) \Mt  - (D+g Q + i\CE ) M - g \Qt Q \)
\end{equation}
A functional integral over the $Q_{\pm}, \Qt_{\pm} $ fields gives the effective
interaction $\CL_{eff} = 
g {\rm Str} ( M_+ \Mt_+ )$
$+
g {\rm Str} ( M_- \Mt_- )$.   If  instead we
first perform
the gaussian functional integral over the $\Up , \Sig$ fields, we obtain
the effective action for the $Q, \Qt$ fields: 
\begin{equation}
\label{sig11} 
S_{Q,\Qt} = {\rm STr} \log \( D - g \Qt - i \CE \) 
+ {\rm STr} \log \( D + g Q + i\CE \)  - g \int \frac{d^2 x}{2\pi}
 ~ {\rm Str} \Qt Q 
\end{equation}
In the above equation, ${\rm STr}$ incorporates the integral $d^2 x$:
For a diagonal functional operator $\CA (x,y) = A(x) \delta (x-y)$ we
define
${\rm STr} \CA = \inv{a^2} \int \frac{d^2 x}{2\pi} {\rm Str} A (x) $,
where $a$ is an ultraviolet cutoff.

The symmetry transformations on $M, \Mt$ follow from 
(\ref{sig6}). 
The lagrangian (\ref{sig10})  then has the following symmetry 
when $\CE = 0$: 
\begin{eqnarray}
\label{symm}
M &\to& \G M \G^\dagger , ~~~~~~~~ \Mt \to \Gt \Mt \Gt^\dagger 
\\ \nonumber 
Q &\to&  \( \G^\dagger \)^{-1} Q \G^{-1} , ~~~~~~~~
\Qt \to  \( \Gt^\dagger \)^{-1} \Qt \Gt^{-1} 
\end{eqnarray}
When $\CE \neq 0$,  the symmetry is as in (\ref{symm})  with 
$G$ an element of $U(1|1)$.

Due to the linear terms in $S_{Q, \Qt}$, the fields $Q , \Qt$ 
develop vacuum expectation values   $\langle Q \rangle$,
 $\langle \Qt \rangle$.  
Let us take the 
vacuum expectation value (VEV) to be 
$\langle (Q_{\pm})_{ab} \rangle = q_{\pm} \delta_{ab}$,
$\langle(\Qt_{\pm})_{ab} \rangle = {\tilde{q}}_{\pm} \delta_{ab}$.  
Minimizing $S_{Q, \Qt}$ with respect to $Q, \Qt$, one finds
that for ${\cal E} \to 0$, 
the VEV's
$q= q_{\pm}$ 
and
${\tilde q}={\tilde q}_{\pm}$
must be solutions to a
set of self-consistent equations which have the solution 
\begin{equation}
 q= \tilde q
= {1 \over a \ g} 
(\exp({4 \pi \over  g}) -1)^{-1/2}
\end{equation}
which for small $g$ goes to zero exponentially as follows: 

\begin{equation}
q = {\tilde q}
\sim
{1 \over a \ g}
\exp(-{2 \pi \over   g})
\end{equation}  
This follows from the saddle point equations
$$
Q=-  \ <x| (D-g {\tilde Q})^{-1} |x>,
\qquad
{\tilde Q}=   \ <x| (D+g Q)^{-1} |x>
$$
which, when written in momentum space,
lead to
$$
{g   {\tilde q} \over 4  \pi }
 \log ( 1 + {1\over  {( a \ g\tilde q)}^2})
= q
$$
(and a similar equation with $q\leftrightarrow {\tilde q}$).

Since the VEV breaks the $GL(1|1; C)$ symmetry
to $U(1|1)$, there are massless Goldstone modes.  These Goldstone
modes as usual can be viewed as a symmetry transformation of the VEV. 
Thus, let us define the fields 
\begin{eqnarray}
\label{cu}
\CU &=&  ( \G^\dagger )^{-1}  \langle Q \rangle \G^{-1} 
=  \left( \matrix { 
q G G^\dagger & 0 \cr 
0 &  q (G G^\dagger )^{-1} \cr 
} \right ) 
=  \left( \matrix{ 
 q U & 0 \cr 
0& q U^{-1} \cr 
} \right) 
\\  \nonumber 
\tilde{\CU} 
&=&  ( \Gt^\dagger )^{-1}  \langle \Qt  \rangle \Gt^{-1} 
=
 \left( \matrix {
  q (G G^\dagger)^{-1} & 0 \cr 
0 & q G G^\dagger  \cr 
} \right) 
=  \left(  \matrix {
 q U^{-1} & 0 \cr 
0 & q U \cr
 } \right) 
\end{eqnarray}
The latter are expressed in terms of $U \equiv  G G^\dagger$. 
An element $G \in GL(1|1; C)$ can be factorized as $G = G_a G_u$ 
where $G_u^\dagger G_u =1$ and $G_a^\dagger = G_a $.  One has  
$U = G_a G_u G_u^\dagger G_a = G_a^2 $, thus the field $U$ lives
on the coset space $GL(1|1;C)/U(1|1)$.      

A  parametrization of the Goldstone modes $U$ 
which manifests the required properties that 
$U$ be invertible and $U^\dagger = U$ is the following:
\begin{eqnarray}
\label{param}
U &=& \left(\matrix{ e^{\phi_1} & 0 \cr 0 & e^{\phi_2}  \cr } \right) 
\left( \matrix{ 1 -  \chi^\dagger \chi & \sqrt{2} \chi \cr 
\sqrt{2} \chi^\dagger & 1-  \chi \chi^\dagger \cr } \right)   
\left(\matrix{ e^{\phi_1} & 0 \cr 0 & e^{\phi_2}  \cr } \right)
\\ \nonumber 
U^{-1}  &=& \left(\matrix{ e^{-\phi_1} & 0 \cr 0 & 
e^{-\phi_2}  \cr } \right) 
\left( \matrix{ 1 -  \chi^\dagger \chi & - \sqrt{2} \chi \cr 
- \sqrt{2} \chi^\dagger & 1-  \chi \chi^\dagger \cr } \right) 
\left(\matrix{ e^{-\phi_1} & 0 \cr 0 & e^{-\phi_2}  \cr } \right)
\end{eqnarray}
The fields $\phi_{1,2}$ are bosonic whereas $\chi, \chi^\dagger$ are
fermionic.  

\def\Str{{\rm Str}}

The low energy $\sigma$-model for the Goldstone modes $U$ is obtained
by substituting $Q, \Qt \to \CU , \tilde{\CU}$ and performing
a derivative expansion using 
$ {\rm STr} \log (A + D) = {\rm STr} \log A + {\rm STr} \log (1 + A^{-1} D)$. 
The zero-th order term in $\CE$ is
\begin{equation}
\label{sigacto}
-\inv{2 a^2  g^2}  
\[ {\rm Str} \( D \CU^{-1} D \CU^{-1} + D \tilde{\CU}^{-1} D \tilde{\CU}^{-1} 
\) \] = - \frac{ q^{-2}}{ a^2 g^2 }  
\Str \d_\mu U^{-1} \d_\mu U 
\end{equation}
Keeping only the linear term in $\CE$, and expressing everything 
in terms of $U$, one finally obtains 
\begin{equation}
\label{sigact}
S=
\int d^2 x \biggl [
-{1\over 8 \pi \lambda^2}
Str
\(\d_\mu U^{-1} \d_\mu U \)
-
{1\over 2\pi}
{\lambda_A\over \lambda^2}
\( \Str U^{-1} \d_\mu U \)^2
-i \CE \lambda_3 ~ \Str (U + U^{-1} ) 
\biggr ]
\end{equation} 
where for small $g $ 
\begin{equation}
\label{gs} 
{1\over \lambda^2}
 = \frac{4}{a^2 g^2  q^2 }\approx  4 \exp( {4 \pi\over g} ) \gg 1
  , ~~~~~ {\lambda}_3 = \frac{2}{a^2 g q}  \approx  {2\over a} 
  \   \exp({ 2 \pi \over g}) 
\end{equation}
As explained below, the $\lambda_A/\lambda^2$ 
term is generated under renormalization.  

In terms of the parametrization (\ref{param}) we find the following
expressions: 
\begin{eqnarray}
\label{sstr}
-\inv{4} \Str \( \d_\mu U^{-1} \d_\mu U \)  &=&  
\bigr \{
\partial_\mu \phi_+ \partial_\mu  \phi_- 
+ \partial_\mu\chi^\dagger \partial_\mu \chi
\bigr \}
+
\chi^\dagger\chi \ \  (\partial_\mu\phi_-)^2
\\
\inv{4} \( \Str U^{-1} \d_\mu U   \)^2  &=& 
\( \d_\mu \phi_-  \)^2 
\\
\Str (U + U^{-1} ) &=&  2 (1 - \chi \chi^\dagger ) \cosh 2 \phi_2 
- 2 (1 - \chi^\dagger \chi ) \cosh 2 \phi_1 
\end{eqnarray}
\begin{equation}
{\rm where} \qquad \qquad
 \phi_{\pm} \equiv (\phi_2 \pm \phi_1)
\end{equation}
(The effect of a  Wess-Zumino topological term is discussed 
below.)

We may study the sigma model as a perturbed conformal
field theory. As an example we will compute
the density of states
for the present model.
Rescaling  the fields $\phi_{1,2} \to  {2\over \lambda}  \, 
 \phi_{1,2}$, 
$\chi \to  {2\over \lambda}
\, \chi$, the sigma model action can 
be written as 
\begin{equation}
\label{R1}
S = S_{CFT} + \int \frac{d^2x}{2\pi} 
\( \frac{\lambda^2}{4} ~   \CO_g + \lambda_A \CO_2   - i \CE \CO_E \) 
\end{equation}
where the conformal field theory has the same field content as
our $gl(1|1)$ model, with action 
\begin{equation}
\label{R2} 
S_{CFT} = \int \frac{d^2x}{8\pi} 
~~ \bigl (
 \( \d_\mu \phi_2 \)^2  - \( \d_\mu \phi_1 \)^2 + \d_\mu \chi^\dagger 
\d_\mu \chi 
\bigr )
\end{equation}
and 
\begin{eqnarray}
\label{R3new}
\CO_g &=&   \inv{4} \chi^\dagger\chi \ (\d_\mu \phi_-)^2 
\\
\CO_2 &=& -  \( \d_\mu \phi_- \)^2 
\\
\CO_E &=& 2 \lambda_3  \[ (1-{\lambda^2 \over 4} \chi \chi^\dagger ) 
\cosh ( \lambda \phi_2 ) 
- (1-{\lambda^2 \over 4} \chi^\dagger \chi ) 
\cosh ( \lambda  \phi_1 ) \] 
\end{eqnarray}

Since $\chi^\dagger\chi$ is a logarithmic operator\cite{Gurarie},  it  mixes
under RG scale transformations with the identity operator, thereby
generating the $\lambda_A $ coupling.

This can also be seen, alternatively, by making the  
following change of variables
$$
\xi^\dagger \equiv e^{\phi_-}\chi^\dagger,
\qquad
\xi \equiv e^{\phi_-}\chi
$$
which yields
\begin{equation}
\label{strgauge}
-\inv{4} \Str \( \d_\mu U^{-1} \d_\mu U \)  =
\bigr \{
\partial_\mu \phi_+ \partial_\mu  \phi_-
+e^{-2\phi_-} \  \partial_\mu\xi^\dagger \partial_\mu \xi
\bigr \}
-
\ \  (\partial_\mu\phi_-) \ \
\partial_\mu [e^{-2\phi_-} \ \xi^\dagger \xi]
\end{equation}
The last term can be eliminated by shifting $\phi_+$,
\begin{equation}
\label{shiftingphiplus}
\phi_+ \to {\Phi_+}\equiv
\phi_+
-e^{-2\phi_-} \ \xi^\dagger\xi
\end{equation}
Let us now also  consider a
topological  Wess-Zumino-Witten term  with coupling $k$.
To lowest (cubic) order this term is easily computed with the
result
\begin{equation}
\label{WZWDEF}
S_{WZW}=
{i k \over 24 \pi }  \int_{M_3} d^3 x
\  \epsilon^{ijk} \
\Str
\biggl \{
U^{-1}\partial_i U \
[
U^{-1}\partial_j U, \
U^{-1}\partial_k U]
\biggr \}
=
\end{equation}
$$=
{i k \over 2 \pi}  \int d^2 x
\
(\partial_\mu \phi_-)
\epsilon^{\mu\nu}
[
(\partial_\nu \xi^\dagger) \xi
-
\xi^\dagger
(\partial_\nu \xi)
]+
{\rm quartic \ and \ higher \  terms}
$$
where, as usual,  $M_3$ is a three-dimensional manifold whose
boundary is the two-dimensional space of interest
and $\epsilon^{\mu\nu}$ the  two-dimensional  antisymmetric tensor.

The  sigma model including the WZW term can be solved using the Lagrange multiplier method of section IV.
In order to see this, note that the fundamental matrix field is of the form
\begin{equation}
\label{Ufactorization}
U= e^{{ \Phi_+}} \ V
\end{equation}
where $V$ does not depend on ${ \Phi_+}$. This implies that the commutator
in Eq.(\ref{WZWDEF}) does not contain any $\Phi_+$-dependence.
Since the supertrace of a commutator vanishes, the WZW term does not
contain any $ \Phi_+$ dependence either\footnote{ Note that this argument
is equally valid for the $GL(N|N)/U(N|N)$ generalization of section VI B
below.}.
Therefore, as  for the current-current perturbation of section IV,
$ \Phi_+$ acts as a Lagrange multiplier, and we can use the same
steps to solve the present theory.

Making the same rescalings of the fields as above, the
action for the sigma model with WZW term  is
\begin{equation}
\label{SsigmaWZW}
S=
S_{CFT}+
\int \frac{d^2x}{2\pi}
\(
 {\lambda \over 2}  \ {\cal O}_g
-
{1 \over 2} k \lambda^3 \  {\cal O}_{WZW}
+ \lambda_A \CO_2  -i \CE \CO_E \) + ...
\end{equation}
where\footnote{No confusion should arise 
between the fields $\Phi_{\pm}$, $\Phi_{1,2}$
defined here and those  of Eq. (\ref{sig1},\ref{sig2})
and in Appendix A,B.}
\begin{eqnarray}
\label{DEFPhipm}
 { \Phi}_{+} &\equiv& ({ \Phi}_2 + { \Phi}_1),
\qquad
 \Phi_{-}\equiv \phi_{-} \equiv  ({ \Phi}_2 - { \Phi}_1)
\end{eqnarray}
and
only terms of linear order in $\Phi_-$ have been 
written\footnote{As in section IV, the term linear in
$\Phi_-$ will be sufficient in the following.}.
Here $S_{CFT}$ is given as before by Eq.(\ref{R2}) 
upon replacing, $\phi_{\pm}\to \Phi_{\pm}$, 
$\chi\to \xi$,
and
\begin{eqnarray}
\label{R3}
\CO_g &=&  
- \inv{2}  \Phi_- (\d_\mu \xi^\dagger \partial_\mu \xi ) \\
{\cal O}_{WZW} &=& \inv{2} \Phi_-  \ i \epsilon^{\mu\nu}
(\d_\mu \xi^\dagger \partial_\nu \xi) \\
{\cal O}_2 &=& -(\partial_{\mu}\Phi_-)^2 \\
\CO_E &=&
 2 \lambda_3 
\[
\cosh ( \lambda { \Phi}_2 )
-\cosh (  \lambda  {  \Phi}_1 )
-{\lambda^2 \over 4}  \xi^\dagger\xi 
\( 
e^{ \lambda { \Phi}_1}+
e^{\lambda { \Phi}_2}
\)
\].
\end{eqnarray}

We proceed with the 1-loop R.G.  equations. One has the OPE's:
\begin{equation}
\label{R6}
\CO_g (z, \zbar) \CO_g (0) \sim { -1 \over 4z\zbar}  \CO_2 (0) + ....
\end{equation}
\begin{equation}
\label{R6WZW}
{\cal O}_{WZW} (z, \zbar) {\cal O}_{WZW} (0)
\sim { +1 \over 4z\zbar}  \CO_2 (0) + ....
\end{equation}
Thus, as in section IV, to lowest order the $\beta$eta-functions are
\begin{equation}
\label{R7}
\beta_{\lambda } = 0 , ~~~~~~\beta_{k} =0,
 ~~~~~~\beta_{\lambda_A} =   {\lambda^2 \over  16}
[1- (k \lambda^2 )^2] +...
\end{equation}
The sigma model coupling constant  $\lambda$,
 and the WZW coupling $k$  generate
 $\lambda_A $ terms (denoted $g_A$ in section IV) of opposite signs.
The 1-loop R.G. equations have
the same structure as those found for the $gl(1|1)$ current-current
perturbation model in
section IV,
where $\CO_2$ is analogous to $\CO_A$.
In the presence of a WZW term, the sigma model becomes
the conformal $gl(1|1)$ WZW model   at a particular
 finite value
 of the sigma model coupling $ 1/ \lambda^2 = k$.

We may solve the theory in Eq.(\ref{SsigmaWZW})
exactly using the Lagrange multiplier method of section IV A.
In order to compute correlators of $ \Phi_+$ one needs
$$
4 \pi {\delta  S\over \delta \Phi_-(z,{\bar z})}_{| \Phi_-=0}
=
  {\lambda \over 2}
\biggl (
- (\d_\mu \xi^\dagger \partial_\mu \xi )
+
(  k \lambda^2)
i \epsilon^{\mu\nu}
(\d_\mu \xi^\dagger \partial_\nu \xi)
\biggr ) =
$$
$$
=
-  \lambda 
\biggl (
[1-  (  k \lambda^2 )] \partial_z \xi^\dagger \partial_{\bar z} \xi
+[1+ (k \lambda^2)] \partial_{\bar z} \xi^\dagger \partial_z \xi
\biggr )
$$
As in section IV A this gives the two point 
function of ${ \Phi_+}$ fields:
$$
<\partial_z\partial_{\bar z}{ \Phi}_+(z,{\bar z})
\
\partial_w\partial_{\bar w}{ \Phi}_+(w,{\bar w})>_{\lambda_A=0}
=
-{\lambda^2\over 2}
[ 1-(  k \lambda^2)^2]
\ {1\over
(z-w)^2 ({\bar z} - {\bar w})^2 } + \delta-{\rm functions}
$$
Furthermore, as in Eq.(\ref{eIxxxvi}) we find
$$
<{ \Phi}_+(z,{\bar z}) \ \Phi_-(0)>
=-2\log (z{\bar z}), 
\qquad 
<\Phi_-(z,{\bar z}) \ \Phi_-(0)> =0
$$
As in section IV A, this yields the correlation functions
\begin{eqnarray}
\label{phiphisigma}
<{ \Phi}_1(z, {\bar z})
{ \Phi}_1(0)>
&=&
(1- 4 \lambda_A) 
\log (z {\bar z}/a^2)
-
{ \lambda^2 \over 16}
[ 1- ( k \lambda^2)^2] 
\log^2 (z {\bar z}/a^2) \\
<{ \Phi}_2(z, {\bar z})
{ \Phi}_2(0)>
&=&
-(1+4\lambda_A) 
\log (z {\bar z}/a^2)
-
{\lambda^2 \over 16}
[ 1- ( k \lambda^2)^2] 
\log^2 (z {\bar z}/a^2)
\end{eqnarray}
as well as the {\it exact}  beta function (identical
to the 1-loop result):
\begin{equation}
\label{betasigma}
{d \lambda_A \over dl} = 
\beta_{\lambda_A} =
 {\lambda^2 \over 16}
 [ 1 - (k \lambda^2)^2 ]
\end{equation}

We now proceed with the calculation of the density of states.
The most relevant operator in $\CO_E$ is $\cosh ( \lambda{ \Phi}_2)$,
whose
 anomalous dimension is 
\begin{equation}
\label{R10} 
\Gamma (\lambda_A) = 
 {\rm dim} \( \cosh (  \lambda  { \Phi}_2 ) \) 
= -  \lambda^2 (1+4 \lambda_A ).  
\end{equation}
The analysis of section V applies to the density of
states with $\lambda_A$ playing the role of $g_A$,
yielding
\begin{equation}
\label{R11sigma}
\bar{\rho (E)}  /\bar{\rho (E_R)}= 
{E_R \over E}
 \exp \( -2 [\lambda_A (E_R)-\lambda_A (E)]
 / \beta_{\lambda_A} \) 
\end{equation}
(valid since $\beta_{\lambda_A}$ is independent of $\lambda_A$).
In the limit $ E\to 0$ this gives a divergent 
(but  integrable) density of states
$$
\bar{\rho}(E) \sim 
{1\over E}
\exp \bigl \{
- {4 \sqrt{2}\over \lambda^2 \sqrt{1 - (k\lambda^2)^2} }
\sqrt{ \log (E_R/E)} \bigr \}
$$
in agreement with the perturbative
 result obtained by Gade\cite{Gade}
using replicas when $k=0$, i.e. in the absence of the WZW term.

We end this section by commenting on the connection with
the scale invariant $PSL(1|1)$ sigma model.
 Eliminating the two bosonic coordinates $\phi_\pm$
the fundamental  $PSL(1|1)$ sigma model field becomes
\begin{equation}
\label{pslV}
V=
\pmatrix{
(1 -\chi^\dagger\chi )
& \sqrt{2} \chi \cr
\sqrt{2} \chi^\dagger & (1
 + \chi^\dagger\chi) \cr}
\end{equation}
whose dynamics is governed by the quadratic action obtained from
$S_{CFT}$ of Eq.(\ref{R2}) by eliminating $\phi_1, \phi_2$.
In  subsection C we briefly discuss the generalization
to the less trivial  $PSL(N|N)$ case.

\subsection{ $GL(1|1)/U(1|1)$ Sigma Model
as a Current-Current Perturbation}

Upon comparing the correlation functions of the
sigma model (Eq.(\ref{phiphisigma}) ) with
those of the $N=1$-species XY model (Eq.(
\ref{eIxxxx},
\ref{eIxxxxb})),
 which is a current-current perturbation,
one finds that the two models have identical correlation
functions if one identifies the coupling constants
as follows:
\begin{eqnarray}
\label{currcurrsigma}
2 \lambda_A &=& \zeta^2 g_A \\
{1\over 4} \lambda^2
[1-(k\lambda^2)^2]
&=&
(2 g )^2 \zeta^6,
\qquad \bigr ( {\rm where} \ \  \zeta^2 =1/(1+2g) \bigr )
\end{eqnarray}
 The identity of the sigma model with the current-current
perturbation of the conformal current algebra 
is likely a special case of a more general statement (see next subsection).

\subsection{ $GL(N|N)/U(N|N)$  Sigma Model }

The generalization of the sigma model of
the previous section to $GL(N|N)$ is immediate.
We start with  a theory of $2N$ bosons and $2N$
fermions (in the previous section $N=1$).
The construction of the previous section
provides us in the standard way with a
field variable on a manifold that we denote by  $U\in GL(N|N;C)/U(N|N)$.
Denoting the  elements of the Super-Lie
algebra $gl(N|N)$ by
$J_{\pm}$ and $J_A$, 
a parametrization of the fundamental matrix field, say of the form\footnote{
This follows as in the $N=1$ case from the
decomposition $G=G_a G_u$, see  below Eq.(\ref{cu}).}
$$
U[X_{\pm}, X_A]
\equiv
e^{\{ X_+ J_+ \}}
 \ \exp\{ X_- J_-
+
\sum_{A\not = \pm}
X_A J_A
\}
$$
yields an action generalizing
Eq.(\ref{sstr}). The $PSL(N|N)$ version is obtained
by setting the coordinates $X_\pm$ to zero.
This sigma model in the presence of a WZW term with coupling $k$
becomes the $gl(N|N)$ WZW conformal field theory at a finite
value $\lambda^2 = 1/k$ of the sigma model coupling constant. 
The latter is a 
$gl(N|N)$ current algebra at level $k$.
 Generalizing the observation made in the preceding
subsection for the $N=1$ case,  we suggest that the 
line of fixed  points arising from
the current-current perturbation of the $PSL(N|N)$ 
variant of the current algebra analyzed in section IV
  is an alternative and `dual' description
of the $PSL(N|N)$ sigma model with  WZW term,  as the
sigma model coupling is varied away from the scale invariant
point.
We plan on reporting in more detail on this connection
in a subsequent publication.

\bigskip

\section{ Conclusions}
In this paper we have studied current-current perturbations
of $gl(N|N)$ super-current algebras.  The existence of two
quadratic Casimir invariants leads to a `nearly conformal'
theory for any value of the coupling. 
Current algebra techniques allowed us to compute
exactly the correlation functions of all
current operators in the perturbed theory,
as well as the exact $\beta$etafunctions (in a particular
RG scheme).
The current correlators are conformal except  for correlation
functions involving  the currents corresponding to
the trace and the supertrace of the super-Lie algebra. 
In  the case $N=1$ we have computed the correlation functions
of {\it all} fields. 
We have applied this to two different, but related disordered
models, the  2D random field XY Statistical Mechanics model, 
and the delocalization transition of electrons
hopping on a 2D square lattice with $\pi$-flux per plaquette
and real hopping amplitudes 
(HWK,  \cite{hatsugai}). 
Both theories can be formulated in terms of the 2D Dirac
hamiltonian subject to random real mass, imaginary scalar
and imaginary vector potentials. This hamiltonian is invariant
under two discrete symmetries, charge-conjugation (particle-hole)
and time-reversal, in every realization of disorder.
For the random XY model we have computed all correlation
functions involving 1st-moment averages (described
by the `1-species theory'), as well as certain correlation
functions 
involving Nth-moment averages
(namely those involving the $gl(N|N)$ currents). In particular,
  we have computed
the  correlation function  of the 
`Edwards-Anderson order parameter',
which is scale invariant and of scaling dimension $x=2$,
in the 2nd moment ($N=2$-species) theory.
For the random hopping model of Hatsugai et al.
 we have obtained the
density of states which shows a divergence at zero energy.
In section IV D we derived the 1-loop RG equations for
a generalized  random XY model: When formulated in terms
of a random 2D Dirac Hamiltonian, the latter still exhibits
 particle-hole symmetry, but lacks time-reversal symmetry,
as opposed to that of  HWK.  The delocalization
transition exhibited by this more general random Dirac hamiltonian
is in the Gade-Wegner universality class, describing
hopping of electrons on a 2D square lattice 
with general complex hopping amplitudes.
In section VI we have
derived a SUSY sigma model from the underlying random
Dirac hamiltonian, found to be defined on
 a  target manifold
which we denote by $GL(N|N;C)/U(N|N)$. We have solved
this sigma model exactly for $N=1$ (including also 
a WZW term), and shown that it is identical to
a current-current perturbation of a $GL(1|1)$ super-Lie
current algebra. The density of states
is obtained from the sigma model.

\section{Acknowledgments} We would like to thank M P A Fisher 
for early discussions
about the replica version of the model by Hatsugai, Wen and Kohmoto.
We also benefited
from conversations with D. Bernard.  This work was supported
in part by the foundation FOM of the Netherlands (S.G.),
by the National Science Foundation, in particular through the
National Young Investigator program (A.LeC.), 
and in part by the A P Sloan Foundation (A.W.W.L).

\appendix

\section{SUSY Approach to the Random Field XY-Model}

In this Appendix we show that the  SUSY approach to
the random field XY model leads to the $N=1$ species
action discussed in the bulk of the paper.

The non-hermitian hamiltonian relevant for the 2D random field XY model
is
$$
H_2=(-i\partial_x + i A_x)\sigma_1
+
(-i\partial_y +  iA_y)\sigma_2
+
M\sigma_3
-
iV {\bf 1}_2
$$
In order to be able to integrate over bosonic variables, we need to consider
the hermitian hamiltonian
\begin{eqnarray}
    H_4
&\equiv&
\pmatrix{0 & H_2 \cr
  H^\dagger_2 &  0  \cr} =
\\ 
 &=&
-i\partial_x \sigma_1\otimes\tau_1
-i\partial_y \sigma_2\otimes\tau_1
+  A_x\sigma_1\otimes\tau_2
+  A_y\sigma_2\otimes\tau_2
+ M\sigma_3\otimes \tau_1
- V{\bf 1}_2\otimes \tau_2
\end{eqnarray}
In the case of interest where all potentials $A_x, A_y, M, V$ are real,
$H_4$ is time reversal invariant, which means that
$H_4$ can be brought into real, symmetric form.
Explicitly, this can be achieved by conjugating
$H_4$  with the matrix
$$
U=U_{\sigma}\otimes U_{\tau},
\qquad {\rm where} \qquad
U_{\sigma}\equiv (1-i\sigma_1)/\sqrt{2}, 
\ \ 
U_{\tau}\equiv
(1-i\tau_3)/\sqrt{2} 
$$
yielding the  real symmetric  hamiltonian
\begin{equation}
\label{realsymm}
H_4^s\equiv U^\dagger H_4 U
=
i \partial_x ( \sigma_1\otimes \tau_2 )
+i\partial_y ( \sigma_3\otimes \tau_2 )
-V ( 1\otimes \tau_1)  - M ( \sigma_2\otimes \tau_2)
\end{equation}

For integration over fermionic variables an antisymmetric form is needed.
This can be obtained by conjugating by
$$
U'\equiv
U_{\sigma}\otimes{\bf 1}_2
$$
\begin{equation}
\label{Hfourantisymm}
H_4^a
\equiv
U'^\dagger H_4 U' 
=
i \partial_x ( \sigma_1\otimes \tau_1)
+i\partial_y ( \sigma_3\otimes \tau_1)
+V (1\otimes \tau_2) - M (\sigma_2\otimes \tau_1)
\end{equation}

All correlation functions relevant for the XY model can be
obtained from an action constructed 
with a {\it (real) 4-component fermion field}
$\chi$, 
and a {\it real 4-component boson field} $\varphi$, defined as follows:
$$
{\bf \chi} \equiv \pmatrix{  \chi_1 \cr\chi_2 \cr\chi_3 \cr\chi_4 \cr}
=\pmatrix{\chi_+ \cr \chi_-\cr}
,
\qquad \qquad
\varphi \equiv \pmatrix{ \varphi_1 \cr\varphi_2 \cr\varphi_3 \cr\varphi_4 \cr}=
\pmatrix{\varphi_+ \cr \varphi_- \cr},
$$
with Lagrangian (density)
$$
{\cal L}
\equiv 
{\cal L}_f
+{\cal L}_b, \qquad\qquad
{\cal L}_f
\equiv
i \chi^t U'^\dagger H_4 U' \chi,
\qquad \qquad
{\cal L}_b
\equiv
i \varphi^t U^\dagger H_4 U \varphi
$$
It will prove convenient to introduce  $4$-component complex fields and their
adjoints by
$$
{\rm Fermions:} \qquad
 \Psi \equiv \pmatrix{\Psi_+\cr \Psi_-\cr} \equiv
U' \ {\bf \chi},
\qquad {\bar \Psi}\equiv \pmatrix{{\bar \Psi_+}, & {\bar \Psi_-}\cr} \equiv
{\bf \chi}^t U'^\dagger
$$
$$
{\rm Bosons:}\qquad
\Phi \equiv
\pmatrix{\Phi_+\cr \Phi_-\cr} \equiv
U {\bf\varphi},
\qquad
\Phi^\dagger
\equiv
\pmatrix{\Phi_+^\dagger, & \Phi_-^\dagger \cr} \equiv
{\bf\varphi}^t U^\dagger
$$
so that\footnote{
Note that the latter equalities involving $\Psi_{\pm }$, $\Phi_{\pm}$
arise because 
$H_4^s$ is symmetric, and because
$U'$ and $U$ do not mix the $2\times 2$ blocks involving $H_2$.}
$$
{\cal L}_f
=
i {\bar \Psi} H_4 \Psi=
 2 i {\bar \Psi}_+ H_2 \Psi_-=
 2  i {\bar \Psi}_- H^\dagger_2 \Psi_+, 
\qquad 
{\cal L}_b
=i \Phi^\dagger  H_4  \Phi
= 2 i \Phi_+^\dagger  H_2  \Phi_-
= 2 i \Phi_-^\dagger  H^\dagger_2  \Phi_+
$$
The so-defined fields are not independent from their adjoints. Rather,
the above  definitions immediately give the following relations:
$$
({\bar \Psi})^t=
U'^* \chi= U'^* U'^\dagger \Psi,
\qquad \qquad
({\Phi}^\dagger)^t=
\Phi^*=
U^* U^\dagger \Phi
$$
Since 
$U'^* U'^\dagger =$ $ U_\sigma^* U_\sigma^\dagger=$
$=-i\sigma_1$ and
$U^* U^\dagger = -\sigma_1\otimes\tau_3$ this implies
$$
\pmatrix{ 
{\bar \psi}_1 \cr 
{\bar \psi}_2 \cr 
{\bar \psi}_3 \cr 
{\bar \psi}_4 \cr 
}
=(-i)
\pmatrix{ 
{\psi}_2 \cr 
{\psi}_1 \cr 
{\psi}_4 \cr 
{\psi}_3 \cr 
},
\qquad \qquad
\pmatrix{ 
\varphi_1^* \cr 
\varphi_2^* \cr 
\varphi_3^* \cr 
\varphi_4^* \cr 
}
=
\pmatrix{ 
-\varphi_2 \cr 
-\varphi_1 \cr 
\varphi_4 \cr 
\varphi_3 \cr 
}
$$
Therefore we may chose
\begin{equation}
\label{appabigpsipm}
{\bar \Psi}_+ =
 \pmatrix{{\bar \psi}_1, & {\bar \psi}_2 \cr}
\equiv
 \pmatrix{\psi^\dagger_R, & \psi^\dagger_L \cr},
\qquad {\rm and} \qquad
\Psi_-
=
\pmatrix{\psi_3 \cr \psi_4\cr}
\equiv
\pmatrix{\psi_L \cr \psi_R\cr}
\end{equation}
as well as
\begin{equation}
\label{appabigphipm}
\Phi_+^\dagger
=
\pmatrix{\varphi_1^*,  & -\varphi_1\cr}
\equiv
\pmatrix{ \beta^\dagger_R, & \beta^\dagger_L\cr},
\qquad {\rm and} \qquad
\Phi_-
=
\pmatrix{ \varphi^*_4 \cr \varphi_4 \cr}
\equiv
\pmatrix{ \beta_L \cr\beta_R \cr}
\end{equation}
as our independent  4  fermionic and 4 (real)  bosonic
integration variables.
The resulting supersymmetric action
$$
{\cal L}_{SUSY}=
i 
 \pmatrix{\psi^\dagger_R, & \psi^\dagger_L \cr}
H_2
\pmatrix{\psi_L \cr \psi_R\cr}
+
i
\pmatrix{ \beta^\dagger_R, & \beta^\dagger_L\cr}
H_2
\pmatrix{ \beta_L \cr\beta_R \cr}
$$
is the $N=1$-species theory discussed in the bulk of the paper.

\section{The Most General Particle-Hole Symmetric Dirac Hamiltonian}

In this appendix we discuss in some detail the  SUSY invariant action
for the most general random particle-hole symmetric,
but not necessarily time-reversal symmetric  Dirac
Hamiltonian.  We start from the hermitian quantum mechanical 
Dirac Hamiltonian $H_4$ defined by
\begin{equation}
\label{hfourAppB}
    H_4
\equiv
\pmatrix{-i\epsilon {\bf 1}_2 & H_2 \cr
H^\dagger_2 &  -i \epsilon { \bf 1}_2 \cr}
\end{equation}
where
$$
H_2 \equiv
(-i\partial_x + i A_x)\sigma_1
+
(-i\partial_y +  iA_y)\sigma_2
+
M\sigma_3
-
iV {\bf 1}_2
$$
(As usual, infinitesimals $\epsilon\to 0^+$ are needed to extract
quantum  mechanical Green's functions).
This model is time reversal invariant if all the potentials  ($A_x, A_y, V, M$)
 are real. In this case
the time reversal operation is implemented by
\begin{equation}
\label{Treversal}
{\cal T } H^*_4 {\cal T} =H_4,
\qquad \qquad {\rm with:} \quad {\cal T }=\sigma_1\otimes\tau_3
\end{equation}

We may consider a more general (non-hermitian) model, where 
we add to any of the potentials an imaginary part, $A_x', A_y', M', V'$.
Note that $A_x', A_y'$ corresponds to a {\it real} vector potential.
The most general hermitian 
hamiltonian $H_4$ incorporating all these potentials is
\begin{eqnarray}
H_4
&=&
-i\partial_x \sigma_1\otimes\tau_1
-i\partial_y \sigma_2\otimes\tau_1
+  A_x\sigma_1\otimes\tau_2
+  A_y\sigma_2\otimes\tau_2
+ M\sigma_3\otimes \tau_1
- V{\bf 1}_2\otimes \tau_2
\\
&~& ~~~~~~
\label{Hfourmostgeneral}
-{A'}_x \sigma_1\otimes\tau_1
-{A'}_y \sigma_2\otimes\tau_1
-M'\sigma_3\otimes\tau_2
+ V' {\bf 1}_2\otimes\tau_1
\end{eqnarray}
In general, this hamiltonian is no longer time reversal invariant\footnote{
the imaginary parts ${A'}_x, {A'}_y, M', V'$ break invariance under 
time reversal, using  ${\cal T}$ defined above.}. However,
it is always invariant under particle-hole transformations
\begin{equation}
\label{Cparticlehole}
{\cal C} \ H_4 \ {\cal C} = - H_4,
\qquad {\rm with} \quad {\cal C} = {\bf 1}\otimes \tau_3
\end{equation}
Time reversal symmetry is recovered
for the special case where  only $A_x, A_y, V', M'$ are non-vanishing. In
this
case the time reversal operation is
\begin{equation}
\label{Tprime}
{\cal T }' H^*_4 {\cal T}' =H_4,
\qquad \qquad {\rm with:} \quad {\cal T }'=\sigma_2\otimes {\bf 1}_2
\end{equation}
These time reversal symmetry properties are summarized in the following table:
\vskip .3cm

\begin{tabular}{cc|c|c|c|}
&non-vanishing  \ potentials    & $A_x, A_y, V, M$   
     & $A_x, A_y, V', M'$     & others \\
\hline
&time reversal     & YES (${\cal T}$)     &  YES (${\cal T}'$)      & NO   \\
\end{tabular}

\vskip 1cm

Let us now return to the most general situation, 
where time reversal symmetry is 
in general absent.
The $4\times 4$ matrix corresponding  to the hamiltonian $H_4$
 has the following block structure:
\begin{equation}
\label{appiH4}
  i \ H_4=
\pmatrix{
\epsilon & 0 &  m - \mu & \partial +A_z +i {A'}_z  \cr
0 & \epsilon & {\bar  \partial} +  A_z^* +i {A'}_z^* & m^*+\mu^* \cr
-m^*+\mu^* & \partial  - A_z +i {A'}_z & \epsilon & 0 \cr
{\bar \partial} - A_z^* +i {A'}_z^*  & -m-\mu & 0 & \epsilon\cr
}
\end{equation}
where
$$
m\equiv (V+iM),
\qquad \qquad
(-\mu) \equiv (M'-iV')
$$
and
$$
A_z\equiv (A_x - i A_y), 
\quad
A_{\bar z}\equiv A_z^*
\qquad \qquad
(-1) {A'}_z\equiv ({A'}_y -i {A'}_x),
\quad
{A'}_{\bar z}\equiv {A'}_z^*
$$
This matrix is manifestly anti-hermitian as it should\footnote{note that  
$\partial \propto [\partial_x-i\partial_y]$,
${\bar \partial} \propto [\partial_x+i\partial_y]$,
so that
$\partial^\dagger=[-\partial_x-i\partial_y]=$
$-{\bar \partial}$.}.

\def\db{\bar{\d}}

\vskip .5cm

Averaged Green's functions of the hamiltonian $H_4$ can be obtained from the
following action.
We  introduce\footnote{The symbols $\beta_{R i}$, $\beta_{L i}$, 
$\beta^\dagger_{R i}$, $\beta^\dagger_{L i}$
will be  used to exhibit the chiral ($L/R$)  nature of the  corresponding
fields (see below).}
a 4-component complex bosonic field  $\Phi$  and its adjoint $\Phi^{\dagger}$,
$$
\Phi \equiv
\pmatrix{ {\phi}_1\cr  {\phi}_2 \cr {\phi}_3 \cr {\phi}_4\cr}
\equiv
\pmatrix{
\beta^{\dagger}_{L2} 
\cr
\beta^{\dagger}_{R2}
 \cr
\beta_{L1} 
\cr
\beta_{R1}
 \cr
}
$$
\begin{equation}
\label{DEFbetagamma}
\Phi^{\dagger}
\equiv
\pmatrix{ {\phi}^*_1, & {\phi}^*_2, & {\phi}^*_3, & {\phi}^*_4\cr}
\equiv
\pmatrix{
{\beta}^{\dagger}_{R1}, &
 {\beta}^{\dagger}_{L1}, &
-{\beta}_{R2}, &
-{\beta}_{L2}\cr
}
\end{equation}
as well as 
a 4-component (complex) fermionic field $\Psi$ and $\Psi^\dagger$:
$$
\Psi
\equiv
\pmatrix{ {\psi}_1\cr  {\psi}_2 \cr {\psi}_3 \cr {\psi}_4\cr}
\equiv
\pmatrix{
\psi^{\dagger}_{L2} \cr
\psi^{\dagger}_{R2} \cr
\psi_{L1} \cr
\psi_{R1} \cr
}
$$
\begin{equation}
\label{DEFpsiLR}
\Psi^{\dagger}
\equiv
\pmatrix{ 
{\psi}^*_1, & {\psi}^*_2, & {\psi}^*_3, & { \psi}^*_4
\cr}
\equiv
\pmatrix{
{\psi}^{\dagger}_{R1}, &
{\psi}^{\dagger}_{L1}, &
{\psi}_{R2}, &
{\psi}_{L2}\cr
}
\end{equation}
The SUSY lagrangian (density) reads,  
in the absence of the vector potentials:
$$
 {\cal L}_{susy} =
\Psi^{\dagger} i H_4 \Psi +
\Phi^{\dagger} \ i H_4 \ \Phi  = 
$$
$$
=\biggl [
\biggl (
{\psi}^*_1 \partial \psi_4 + {\psi}^*_4 {\bar \partial} \psi_1
\biggr )
+
\biggl (
 \phi^*_1 \partial \phi_4 + \phi^*_4 {\bar \partial} \phi_1
\biggr )
\biggr ]
-
\biggl [
\biggl (
{\psi}^*_3 \partial (- \psi_2) + (-{\psi}^*_2) {\bar \partial} \psi_3
\biggr )
+
\biggl (
\phi^*_3 \partial (- \phi_2) + (- \phi^*_2) {\bar \partial} \phi_3
\biggr )
\biggr ]+
$$
$$
+\biggl [
m_+
\biggl (
{\psi}^*_1 \psi_3 + {\phi}^*_1 \phi_3
\biggr )
+m_-
\biggl (
 {\psi}^*_4 (-\psi_2) + {\phi}^*_4 (-\phi_2)
\biggr )
\biggr ]
-
\biggl [
m_+^*
\biggl (
 {\psi}^*_3 \psi_1 + \phi^*_3 \phi_1
\biggr )
+
m_-^*
\biggl (
(-{\psi}^*_2) \psi_4
+
(-\phi^*_2) \phi_4
\biggr )
\biggr ]+
$$
\begin{equation}
\label{appsusyactionUpsilon}
+\epsilon
\biggl [
{\psi}^*_1 \psi_1
+ \phi^*_1 \phi_1
+ (- {\psi}^*_2) (-  \psi_2)
+ (- \phi^*_2) (-  \phi_2)
+ {\psi}^*_3 \psi_3
+ \phi^*_3 \phi_3
+ {\psi}^*_4 \psi_4
+ \phi^*_4 \phi_4
\biggr ]  
\end{equation}
where 
$$m_+\equiv (m+\mu), 
\qquad \qquad
m_-\equiv (m-\mu).
$$
This may also be expressed in terms of the $R$- and $L$-moving fields, defined
above:
$$
{\cal L}_{susy}
=
$$
$$=
\sum_a (  \psi^\dagger_{Ra} \d \psi_{Ra} + \beta^\dagger_{Ra} \d 
\beta_{Ra} + \psi^\dagger_{La} \db \psi_{La} + \beta^\dagger_{La}
\db \beta_{La} )+
$$
$$
+
m \( \sum_a  (\psi^\dagger_{Ra} \psi_{La} + \beta^\dagger_{Ra} \beta_{La} )
\)
+
m^* \( \sum_a  (\psi^\dagger_{La} \psi_{Ra} + \beta^\dagger_{La} \beta_{Ra} )
\) +
$$
$$
+
\mu
\( 
( {\psi}^{\dagger}_{R1} \psi_{L1} +\beta^{\dagger}_{R1}\beta_{L1}  )
-
( {\psi}^{\dagger}_{R2} \psi_{L2} + \beta^{\dagger}_{R2}\beta_{L2})
\) 
-
\mu^*
\(
({\psi}^{\dagger}_{L1} \psi_{R1} +\beta^{\dagger}_{L1}\beta_{R1} )
- ( {\psi}^{\dagger}_{L2} \psi_{R2} + \beta^{\dagger}_{L2}\beta_{R2}  )
\)+
$$
\begin{equation}
\label{appsusyactionLR}
+
\epsilon
\(
 {\psi}^{\dagger}_{R1} \psi^{\dagger}_{L2}
-
{\psi}^{\dagger}_{R2}
\psi^{\dagger}_{L1}
+
{\psi}_{L2}
\psi_{R1}
-
{\psi}_{L1}
\psi_{R2} 
+
  \beta^{\dagger}_{R1} \beta^{\dagger}_{L2} +
\beta^{\dagger}_{R2} \beta^{\dagger}_{L1}
-
 \beta_{R1}\beta_{L2} + \beta_{R2}\beta_{L1}
\) 
\end{equation}

The vector potentials give rise to an additional contribution to
the action of the form:
\begin{eqnarray}
\nonumber
{\cal L}_{susy}^A
&\equiv&
 A_z
\( \sum_a 
(\psi^\dagger_{Ra}\psi_{Ra}
+
\beta^\dagger_{Ra}\beta_{Ra})
\)
+
 A_{\bar z}
\(\sum_a 
(\psi^\dagger_{La}\psi_{La}
+
\beta^\dagger_{La}\beta_{La})
\)
\\ \nonumber
&+&i  {A'}_z
\( 
\psi^\dagger_{R1}\psi_{R1}
+
\beta^\dagger_{R1}\beta_{R1}
-
\psi^\dagger_{R2}\psi_{R2}
-
\beta^\dagger_{R2}\beta_{R2}
\)
+ i  {A'}_{\bar z}
\(
\psi^\dagger_{L1}\psi_{L1}
+
\beta^\dagger_{L1}\beta_{L1}
-
\psi^\dagger_{L2}\psi_{L2}
-
\beta^\dagger_{L2}\beta_{L2}
\)=
\\ \nonumber
&=&
A_z
\biggl
[
({{\bar J}'}_{11}-{\bar J}_{11}) 
+
({{\bar J}'}_{22}-{\bar J}_{22}) 
\bigr ]
+
A_{\bar z}
[
({J'}_{11}-J_{11}) 
+
({J'}_{22}-J_{22}) 
\bigr ]+
\\ 
\label{vectorpotentials}
&+&i{A'}_z
\biggl
[
({{\bar J}'}_{11}-{\bar J}_{11}) 
-
({{\bar J}'}_{22}-{\bar J}_{22}) 
\bigr ]
+
i {A'}_{\bar z}
[
({J'}_{11}-J_{11}) 
-
({J'}_{22}-J_{22}) 
\bigr ]
\end{eqnarray}

\section{ 1-loop RG Equations 
for the Generalized Random XY Model (Broken Time Reversal Symmetry)}

In this Appendix we analyze the 1-loop RG equations for the random
XY model without time  reversal symmetry.

Let us first consider the RG equations for the couplings $g_1=g_m-g_\mu $ and 
$g_2=g_m+g_\mu $, 
defined in Eq.'s (\ref{Lbarsusy},  \ref{DEFgonegtwo}).
The most general form\footnote{ Note that
we have omitted terms on the r.h.s. which are quadratic in the
vector potential couplings, i.e. of the form  $g^A_i g^A_j$.
The operators  coupling to  $g^A_i$ are of the form $J_{aa}$ and ${J'}_{bb}$
(see Eq.(\ref{LbarA})). Since there are no poles in the OPE of such
operators with each other (see Eq.(\ref{n5})) they do not
generate 1-loop RG flows.}
 of the 1-loop RG is:
$$
{d g_1\over dl}
=
a_1 g_1^2 + b_1 g_1 g_2 + c_1 g_2^2+
 ( d^1_{11} g_1 + d^1_{21} g_2) g_1^A
+ ( d^1_{12} g_1 + d^1_{22} g_2) g_2^A
$$
$$
{d g_2\over dl}
=
a_2 g_1^2 + b_2 g_1 g_2 + c_2 g_2^2+
 ( d^2_{11} g_1 + d^2_{21} g_2) g_1^A
+ ( d^2_{12} g_1 + d^2_{22} g_2) g_2^A
$$
We proceed in the following steps, to simplify these equations:

\vskip .3cm

\noindent \underbar{\it Step 1:}  We know that both RG equations vanish
when $g_2=g_2^A=0$ describing two decoupled 1-species theories.
This gives
$a_1=d^1_{11}=a_2=d^2_{11}=0$.

\vskip .3cm

\noindent \underbar{\it Step 2:}  Adding and subtracting the two equations:

$$
{ d (g_1+g_2)\over d l}
=2{d g_m\over d l}
=
(b_1+b_2) g_1 g_2 + (c_1+c_2) g_2^2
+
(d^1_{21}+d^2_{21}) g_2 g_1^A + 
[ (d^1_{12}+d^2_{12})g_1 + (d^1_{22}+d^2_{22})g_2 ]g_2^A
$$

$$
{ d (g_1-g_2)\over d l}
=(-2){d g_\mu\over d l}
=
(b_1-b_2) g_1 g_2 + (c_1-c_2) g_2^2
+
(d^1_{21}-d^2_{21}) g_2 g_1^A + 
[ (d^1_{12}-d^2_{12})g_1 + (d^1_{22}-d^2_{22})g_2 ]g_2^A
$$

Setting $g_\mu=g_{A'}=0$, both equations must vanish identically.
In this case
we have $g_1=g_2=g_m$ and $g^A_1=g^A_2=g_A$, which yields, when inserted into
the RG equations:
$$
b_1 + c_1  =
d^1_{21}  +  d^1_{12}+ d^1_{22}=
b_2 + c_2  =
d^2_{21}  +  d^2_{12}+ d^2_{22}=0
$$

\vskip .3cm
\noindent \underbar{\it Step 3:}  Finally, setting
$g_m=g_{A'}=0$, again both equations must vanish.
In this case
we have $g_1=-g_\mu, g_2=g_\mu$ and 
$g^A_1=g^A_2=g_A$, which yields, when inserted into
the RG equations:
$$
-b_1 + c_1  =
d^1_{21}  -  d^1_{12}+ d^1_{22}=
-b_2 + c_2  =
d^2_{21}  -  d^2_{12}+ d^2_{22}=0
$$

\vskip 1cm

Combining Steps 1, 2 and 3 we find:

$$
a_1=d^1_{11}=a_2=d^2_{11}=0
$$
$$
b_1=b_2=c_1=c_2=0,
\qquad
d^1_{12}=d^2_{12}=0,
$$
\begin{equation}
\label{steptwothree}
(d^1_{21}+ d^1_{22})=
(d^2_{21}+ d^2_{22})=
0
\end{equation}

\vskip 1cm

The 1-loop RG equation thus simplifies to
$$
{d g_1\over dl}
=
  ( d^1_{21} g_1^A
+   d^1_{22}  g_2^A ) g_2 
$$
$$
{d g_2\over dl}
=
  ( d^2_{21}  g_1^A
+  d^2_{22}  g_2^A) g_2
$$

Using Eq.(\ref{DEFgAgAprime})
this becomes
$$
{d g_1\over dl}
=
 - ( d^1_{21} - d^1_{22}) 
g_{A'}  g_2 
$$
$$
{d g_2\over dl}
=
 - ( d^2_{21} - d^2_{22})  
g_{A'}  g_2 
$$

Let us now consider explicitly the relevant OPE's: The coupling constant
$g_{A'}$ couples to a left-right bilinear of a {\it Cartan} generator.
Since the simple pole term 
in the OPE of a  Cartan generator with any operator
can only give back the {\it same} operator,
times a number ($=$ the eigenvalue of the
Cartan generator when applied to this operator),
we know that $g_{A'} g_2$ can only generate again
the operator coupling to $g_2$.
Therefore we know that $(d^1_{21}-d^1_{22})=0$
and the RG equations simplifies further to:
$$
{d g_1\over dl} =0
$$
\begin{equation}
\label{dgonedgtwo}
{d g_2\over dl}
=
 - ( d^2_{21} - d^2_{22})  
g_{A'}  g_2 
\end{equation}
(Recall that 
$g_2 = g_m + g_\mu$, from Eq. (\ref{DEFgonegtwo}).)

 \vskip .5cm

The corresponding  RG equations  for the vector potential
couplings are of the form
$$
{d g_A\over dl}
= \alpha_1 g_m^2 + \beta_1 g_m g_\mu + \gamma_1 g_\mu^2
$$
$$
{d g_{A'}\over dl}
= \alpha_2 g_m^2 + \beta_2 g_m g_\mu + \gamma_2 g_\mu^2
$$
Since $g_{A'}$ is not generated when either  $g_\mu=0$  or
when $g_m=0$,  
we have
$\alpha_2=\gamma_2=0$.
In addition, the flow for $g_A$ is the same in both
cases, implying $\alpha_1=\gamma_1$.
Adding and subtracting the resulting equations
$$
{d g_1^A\over d l}
=
{d (g_A- g_{A'})\over d l}
=
\alpha_1 ( g_m^2 + g_\mu^2)
+(\beta_1-\beta_2)g_mg_\mu
$$
$$
{d g_2^A\over d l}
=
{d (g_A+ g_{A'})\over d l}
=
\alpha_1 ( g_m^2 + g_\mu^2)
+(\beta_1+\beta_2)g_mg_\mu
$$
When $g_2=0$ (i.e.: $g_\mu=-g_m$, $g_1=2g_m$) we know that
$$
{d g_1^A\over d l} = \alpha_1 g_1^2,
\qquad
{d g_2^A\over d l} =0
$$

This finally yields the following RG equations 
($\beta_1=0$, $\beta_2=2 \alpha_1$):
\begin{equation}
\label{dgAdgAprime}
{d g_A\over d l}
=
2 [(g_m)^2 + (g_\mu)^2],
\qquad \qquad
{d g_{A'}
\over d l}
=
  4 \ g_m g_\mu
\end{equation}
where we have used that $\alpha_1=2$ [see Eq.(\ref{eIxi})].

\end{document}